\documentclass[reqno]{amsart}

\usepackage[margin=1.2in]{geometry}
\usepackage{enumerate}
\usepackage{enumitem}
\usepackage{amsthm,amsmath,amsfonts,amssymb,amsxtra,appendix,bookmark,euscript,delarray,dsfont}
\usepackage{enumerate}
\usepackage{amssymb}

\usepackage[dvipsnames]{xcolor}

\usepackage{graphicx}
\usepackage{upgreek}
\usepackage[normalem]{ulem}

\usepackage{array}
\usepackage{longtable}
\newcolumntype{L}[1]{>{\raggedright\arraybackslash}p{#1}}
\usepackage{url}
\usepackage{needspace}
\usepackage{verbatim}

\allowdisplaybreaks[1]

\usepackage{tikz}
\usetikzlibrary{decorations.pathreplacing,calc}

\theoremstyle{plain}
\newtheorem{theorem}{Theorem}[section]
\newtheorem{lemma}[theorem]{Lemma}
\newtheorem{remark}[theorem]{Remark}

\newtheorem{proposition}[theorem]{Proposition}

\newtheorem{corollary}[theorem]{Corollary}

\numberwithin{equation}{section}

\renewcommand{\d}{\mathrm{d}}
\newcommand{\dd}{\,\mathrm{d}}

\DeclareMathOperator{\tr}{Tr}

\DeclareMathOperator{\Ran}{Ran}

\renewcommand{\leq}{\leqslant}
\renewcommand{\geq}{\geqslant}
\renewcommand{\ge}{\geqslant}
\renewcommand{\le}{\leqslant}

\def\nn{\nonumber}
\renewcommand{\to}{\rightarrow}
\renewcommand{\phi}{\varphi}

\newcommand{\norm}[1]{\left\lVert #1 \right\rVert}

\renewcommand{\phi}{\varphi}
\newcommand{\ket}[1]{|#1 \rangle}
\newcommand{\bra}[1]{\langle #1 |}
\newcommand{\lt}{\left}
\newcommand{\rt}{\right}
\newcommand{\bangle}[1]{\lt\langle #1\rt\rangle}
\newcommand{\ip}[2]{\bangle{#1, #2}}

\newcommand\R{{\ensuremath {\mathbb R} }}

\newcommand\C{{\ensuremath {\mathbb C} }}

\newcommand\1{{\ensuremath {\mathds 1} }}

\newcommand{\gH}{\mathfrak{H}}

\newcommand{\dG}{\d\Gamma}

\hypersetup{
  colorlinks=true,
  citecolor=Green,
  linkcolor=NavyBlue,
  urlcolor=black
}

\makeatletter
\def\HL@namedref#1#2\ref#3{\hyperref[#3]{#1~\ref*{#3}}}
\DeclareRobustCommand{\Theorem}{\HL@namedref{Theorem}}
\DeclareRobustCommand{\Theorems}{\HL@namedref{Theorems}}
\DeclareRobustCommand{\Lemma}{\HL@namedref{Lemma}}
\DeclareRobustCommand{\Lemmas}{\HL@namedref{Lemmas}}
\DeclareRobustCommand{\Proposition}{\HL@namedref{Proposition}}
\DeclareRobustCommand{\Propositions}{\HL@namedref{Propositions}}
\DeclareRobustCommand{\Corollary}{\HL@namedref{Corollary}}
\DeclareRobustCommand{\Corollaries}{\HL@namedref{Corollaries}}
\DeclareRobustCommand{\Definition}{\HL@namedref{Definition}}
\DeclareRobustCommand{\Definitions}{\HL@namedref{Definitions}}
\DeclareRobustCommand{\Remark}{\HL@namedref{Remark}}
\DeclareRobustCommand{\Remarks}{\HL@namedref{Remarks}}
\DeclareRobustCommand{\Assumption}{\HL@namedref{Assumption}}
\DeclareRobustCommand{\Assumptions}{\HL@namedref{Assumptions}}
\DeclareRobustCommand{\Example}{\HL@namedref{Example}}
\DeclareRobustCommand{\Section}{\HL@namedref{Section}}
\DeclareRobustCommand{\Sections}{\HL@namedref{Sections}}
\DeclareRobustCommand{\Subsection}{\HL@namedref{Subsection}}
\DeclareRobustCommand{\Subsections}{\HL@namedref{Subsections}}
\DeclareRobustCommand{\Appendix}{\HL@namedref{Appendix}}
\DeclareRobustCommand{\Appendices}{\HL@namedref{Appendices}}
\def\HL@namedeqref#1#2\eqref#3{\hyperref[#3]{#1~\textup{(\ref*{#3})}}}
\DeclareRobustCommand{\Equation}{\HL@namedeqref{Equation}}
\DeclareRobustCommand{\Equations}{\HL@namedeqref{Equations}}
\DeclareRobustCommand{\Condition}{\HL@namedeqref{Condition}}
\DeclareRobustCommand{\Conditions}{\HL@namedeqref{Conditions}}
\makeatother

\newcommand{\F}{\mathrm F}

\title{Absence of binding for the helium anion}
\author[H. Liang]{Hao Liang}
\address{School of Mathematical Sciences, Peking University, Beijing, 100871, China}
\email{leunghao@stu.pku.edu.cn}
\date{}

\begin{document}
\begin{abstract}
We study the helium anion in the nonrelativistic Coulomb model with an infinitely massive nucleus and spin-$\frac12$ electrons. We prove the absence of binding, with $E^{\F}(3,2)=E^{\F}(2,2)$ and no three-electron ground state. A decomposition relative to the helium ground state and completion of the square reduce the problem to a one-dimensional variational inequality.
\end{abstract}
\subjclass[2020]{81V45 (Primary) 35P15, 81Q10, 81V74 (Secondary)}
\keywords{Helium anion, absence of binding, ionization conjecture, Coulomb Hamiltonians, ground-state energy.}
\maketitle
\section{Introduction and main result}\label{sec:main}

In atomic units, we describe $N\geq1$ electrons interacting with an infinitely massive nucleus of charge $Z>0$ fixed at the origin by the Coulomb Hamiltonian
\begin{equation}\label{eq:model}
 H_{N,Z}=\sum_{j=1}^N\left(-\frac12\Delta_j-\frac Z{|x_j|}\right)
       +\sum_{1\leq i<j\leq N}\frac1{|x_i-x_j|}.
\end{equation}
The operator is self-adjoint and bounded below on
$$
 \gH_N^{\F}=\bigwedge^N(\mathfrak h\otimes\C^2),
 \qquad \mathfrak h=L^2(\R^3).
$$
We are interested in the ground-state energy of this system
\begin{equation}\label{eq:ground-state-energy}
 E^{\F}(N,Z)
 :=\inf\sigma(H_{N,Z}|_{\gH_N^{\F}})
 =\inf_{\substack{\Psi\in H^1\cap\gH_N^{\F}, \norm{\Psi}_2=1}}
   \ip{\Psi}{H_{N,Z}\Psi}.
\end{equation}
Any normalized minimizer satisfies the Schr\"odinger equation
$
 H_{N,Z}\Psi=E^{\F}(N,Z)\Psi
$
and is called a ground state. Although $E^{\F}(N,Z)$ is finite \cite{LiebSeiringer10}, the existence of minimizers of \eqref{eq:ground-state-energy} does not necessarily hold.

\subsection{Maximum ionization}

A fundamental question in quantum many-body theory is how many electrons a given collection of nuclei can bind. See the surveys \cite{Nam22,Lewin25}. 

Following Lieb~\cite[Theorem~1]{Lieb84}, we say that the system described by $H_{N,Z}$ is \textit{bound} if $H_{N,Z}$ has a normalized ground state in $\gH_N^{\F}$. Otherwise, we say that there is no \textit{binding}. Let $N_c(Z)$\footnote{Some authors use the \textit{binding inequality} $E^{\F}(N,Z)<E^{\F}(N-1,Z)$ to define \textit{binding} and hence $N_c(Z)$. By the HVZ theorem, this inequality implies ground-state existence, so the resulting maximum is at most $N_c(Z)$ as defined here.} denote the largest binding number, namely, $N_c(Z)$ is the positive integer $N$ for which $H_{N,Z}$ has a normalized ground state in $\gH_N^{\F}$. 

Zhislin \cite{Zhislin60} proved that $E^{\F}(N,Z)<E^{\F}(N-1,Z)$ for $N<Z+1$. Using the atomic HVZ theorem \cite{ZhislinSigalov65}:
\begin{equation}\label{eq:energy-monotonicity}
 \sigma_{\mathrm{ess}}(H_{N,Z}|_{\gH_N^{\F}})
 =\big[E^{\F}(N-1,Z),+\infty\big),
\end{equation}
Zhislin's inequality yields a ground state for $N<Z+1$ and hence $N_c(Z)\geq Z$.

Upper bounds on $N_c(Z)$ are considerably more difficult. The atomic ionization conjecture predicts $N_c(Z)\leq Z+C$ for a universal constant $C$. For physical integer nuclear charges, the sharper conjecture is $N_c(Z)\leq Z+1$ \cite{Nam22}. Lieb \cite{Lieb84} proved $N_c(Z)<2Z+1$ by an argument independent of particle statistics. For fermions, Nam \cite{Nam12} refined the argument to obtain $N_c(Z)<1.22Z+3Z^{1/3}$ for $Z\geq1$. Hundertmark, Pattakos, and Schulz \cite{HPS25} further refined the method, showing that for $Z\geq4$, strict binding $E^{\F}(N,Z)<E^{\F}(N-1,Z)$ implies $N<1.1185Z+4Z^{1/3}$. More recently, an AI-generated manuscript released by OpenAI \cite{OpenAI26} claims to prove the uniform bound $N\leq Z+C$ for atoms with $Z\geq1$ under the same strict-binding condition, with $C$ independent of $Z$.

A complementary line of work studies the large-$Z$ limit. Lieb, Sigal, Simon, and Thirring \cite{LSST88} proved the asymptotic neutrality of fermionic atoms, $N_c(Z)/Z\to1$ as $Z\to\infty$. Fefferman and Seco \cite{FS90} obtained quantitative sublinear bounds on the excess charge. See also Seco, Sigal, and Solovej \cite{SSS90}. Fermi statistics are essential for this conclusion, as the corresponding ratio for bosonic atoms converges to a constant strictly larger than one \cite{Solovej90}.

Related ionization problems arise in simplified density-functional and mean-field models. Uniform bounds on the excess charge are known in Thomas--Fermi--von Weizs\"acker theory \cite{BL85}, Hartree--Fock theory \cite{Solovej03}, and atomic Thomas--Fermi--Dirac--von Weizs\"acker theory \cite{FNV18}. More recently, Chen and Wu \cite{ChenWu26} established asymptotic neutrality for the two-center Thomas--Fermi--Dirac--von Weizs\"acker model using a comparison with the neutral Thomas--Fermi density.

\subsection{The critical regime}

Heuristically, Zhislin's lower bound and the sharper atomic ionization conjecture can be understood in terms of the effective charge $Z-(N-1)$ seen by an electron far from the nucleus.
\begin{itemize}
    \item If $N<Z+1$, the Coulomb tail is attractive and favors binding.
    \item If $N>Z+1$, the Coulomb tail is repulsive and favors electron escape.
\end{itemize}
At $N=Z+1$, the residual atom is neutral and the leading $1/r$ term vanishes. Can the neutral atom still bind the additional electron?

By the HVZ theorem, the detachment threshold is $E^{\F}(Z,Z)$ and the following two cases may occur:
$$
 E^{\F}(Z+1,Z)<E^{\F}(Z,Z)
 \qquad\text{or}\qquad
 E^{\F}(Z+1,Z)=E^{\F}(Z,Z).
$$
Strict inequality guarantees a ground state for the anion. At equality, the ground-state energy lies at the bottom of the essential spectrum, and it remains to determine whether this energy is an eigenvalue.

The hydrogen anion exhibits strict binding, $E^{\F}(2,1)<E^{\F}(1,1)$~\cite{SS74}. Numerical studies suggest equality for several heavier anions, including $\mathrm{He}^{-}$, $\mathrm{Be}^{-}$, $\mathrm{N}^{-}$, and $\mathrm{Ne}^{-}$. See~\cite[Section~3]{KNS99} and, for helium,~\cite{FK08}. Thus helium, at $Z=2$, is the first case along $N=Z+1$ in which equality is expected. We prove this equality and show that the helium anion has no ground state.

\begin{theorem}[Absence of binding for the helium anion]\label{thm:main}
The ground-state energies satisfy
$$
 E^{\F}(3,2)=E^{\F}(2,2),
$$
and $H_{3,2}$ has no ground state in $\bigwedge^3(L^2(\R^3)\otimes\C^2)$.
\end{theorem}

We prove the stronger quantitative estimate in Theorem~\ref{thm:main-reduced}, which implies Theorem~\ref{thm:main}.

\begin{remark}
As a result of Theorem~\ref{thm:main} and Lieb's upper-bound, we have $N_c(2)\in\{2,4\}$.
\end{remark}

\begin{remark}
The spin degeneracy $q=2$ is essential here. More generally, one may replace $\C^2$ by $\C^q$ and consider $H_{N,Z}$ on $\bigwedge^N(\mathfrak h\otimes\C^q)$. For $q=1$, variational calculations in the fully spin-polarized sector support strict binding~\cite{WLYD14}. For $q\geq3$, the two- and three-electron energies coincide with their bosonic counterparts, so strict binding follows from Hogreve's result~\cite{Hogreve11}.
\end{remark}

\begin{remark}
The nuclear charge $Z$ can be treated as a positive real parameter, as we already do in the estimates of Sections~\ref{sec:core} and~\ref{sec:square}. The method can be adapted to prove $E^{\F}(3,Z)=E^{\F}(2,Z)$ for $Z\leq2$ sufficiently close to $2$. The right endpoint $Z=2$ is sharp, since Zhislin's theorem~\cite{Zhislin60,ZhislinSigalov65} gives $E^{\F}(3,Z)<E^{\F}(2,Z)$ for every $Z>2$. The proof reflects this change through the term $(2-Z)\ip{f}{|\cdot|^{-1}f}$ in the lower bound of Proposition~\ref{prop:scalar-source-reduction}. This term is nonnegative for $Z\leq2$. For $Z>2$, it becomes an attractive Coulomb term which cannot be absorbed by the kinetic energy.
\end{remark}

After posting this preprint, I learned that Shelley, Yang, and Tartakovsky~\cite{SYT26} had recently independently obtained similar results in their computer-assisted study of maximum ionization for helium and lithium. The two papers take different approaches.

\subsection{Organization of the paper}

The paper is organized as follows. Section~\ref{sec:pre} introduces the notation and preliminary results, and Section~\ref{sec:strategy} outlines the proof strategy. In Sections~\ref{sec:decomposition}--\ref{sec:square}, we establish the decomposition, control the remainder, and complete the square. Sections~\ref{sec:scalar} and~\ref{sec:1-d-variational} reduce the remaining estimate to a one-dimensional variational problem and complete the proof. Appendix~\ref{app:core-inputs} provides the required estimates for the two-electron system.

\subsection*{Acknowledgments} The author thanks Zhenfu Wang for helpful discussions. The author also thanks Yukimi Goto and Marvin Raimund Schulz for their comments on the preprint, and Alec Shelley for sharing the preprint on helium and lithium. This work was partially supported by the National Key R\&D Program of China (Project No.~2024YFA1015500), and the author was partially supported by the NSFC (Grant Nos.~12595282 and 12171009). 

\bigskip

\paragraph{\textbf{AI disclosure.}}
The author thanks ChatGPT-6 Astra for discussions during the preparation of this manuscript. During these discussions, the AI proposed numerous numerical approaches relying on computer-assisted calculations. Although many of these approaches were difficult to interpret, some provided useful inspiration. The author subsequently developed a simpler mathematical argument that avoids most of these computer-assisted methods. In the present manuscript, ChatGPT-6 Astra primarily assisted with deriving the moment estimates in the appendix and with finding suitable test functions for the appendix and Section~\ref{sec:1-d-variational}. AI tools were also used for language editing, checking calculations, literature searches and bibliographic checks. All mathematical statements, proofs and calculations were independently checked and verified by the author, who takes full responsibility for the content of this manuscript.

\section{Preliminaries}\label{sec:pre}

For $n,m\geq0$, let $\gH_{n,m}^{\F}$ denote the subspace of $\gH_{n+m}^{\F}$ with $n$ up spins and $m$ down spins. Then
\begin{equation}
      \gH_N^{\F}=\bigoplus_{n+m=N}\gH_{n,m}^{\F}\cong\bigoplus_{n+m=N}\Big(\bigwedge^n\mathfrak h\otimes\bigwedge^m\mathfrak h\Big).
\end{equation}
We write
$$
 \mathcal K_{n,m}=\bigwedge^n\mathfrak h\otimes\bigwedge^m\mathfrak h.
$$
Under this identification, $H_{n+m,Z}$ acts on $\mathcal K_{n,m}$ as $H_{n+m,Z}^{\mathrm{sc}}$, the same differential expression acting on scalar functions.

\subsection{Fermionic Fock space}

We work with the scalar Hamiltonian $H_{N,Z}^{\mathrm{sc}}$ on $\mathcal K_N$ and use the scalar realization of fermionic Fock space
$$
 \mathcal F_{\mathrm f}:=\bigoplus_{n,m\geq0}\mathcal K_{n,m},
 \qquad \mathcal K_{0,0}=\C,
$$
with vacuum $1\in\mathcal K_{0,0}$. For $f\in\mathfrak h$ and $\psi\in\mathcal K_{n,m}$, the creation operators are defined by
\begin{equation}\begin{split}
 &(a_\uparrow^*(f)\psi)(x_1,\ldots,x_{n+1};y_1,\ldots,y_m)
 \\[-2pt]
 &\qquad=\frac1{\sqrt{n+1}}\sum_{j=1}^{n+1}(-1)^{n+1-j}f(x_j)
 \psi(x_1,\ldots,\widehat{x_j},\ldots,x_{n+1};y_1,\ldots,y_m),
\end{split}\label{eq:creation-up}\end{equation}
\begin{equation}\begin{split}
 &(a_\downarrow^*(f)\psi)(x_1,\ldots,x_n;y_1,\ldots,y_{m+1})
 \\[-2pt]
 &\qquad=\frac{(-1)^n}{\sqrt{m+1}}\sum_{j=1}^{m+1}(-1)^{m+1-j}f(y_j)
 \psi(x_1,\ldots,x_n;y_1,\ldots,\widehat{y_j},\ldots,y_{m+1}).
\end{split}\label{eq:creation-down}\end{equation}

On $\mathcal F_{\mathrm f}$, the annihilation operators are the adjoints $a_\sigma(f)=a_\sigma^*(f)^*$. These operators are bounded and satisfy the canonical anticommutation relations (CAR). We have
\begin{equation}\label{eq:CAR}
 \{a_\sigma(f),a_\tau^*(g)\}
 =\delta_{\sigma\tau}\ip fg \1,\qquad
 \{a_\sigma(f),a_\tau(g)\}=0,\qquad
 \|a_\sigma(f)\|_{\mathrm{op}}=\|a_\sigma^*(f)\|_{\mathrm{op}}=\|f\|_{\mathfrak h}.
\end{equation}
We lift $H_{N,Z}^{\mathrm{sc}}$ to $\mathcal F_{\mathrm f}$ by setting
$$
 \mathbb H:=\bigoplus_{n,m\geq0}
 H_{n+m,Z}^{\mathrm{sc}}\big|_{\mathcal K_{n,m}},
 \qquad H_{0,Z}^{\mathrm{sc}}=0.
$$
For an orthonormal basis $\{e_j\}_j$ of $\mathfrak h$, we have
\begin{align}
 \mathbb H
 &=\sum_{\sigma\in\{\uparrow,\downarrow\}}\sum_{j,k}
 \langle e_j,h_Ze_k\rangle
 a_\sigma^*(e_j)a_\sigma(e_k)\nn\\
 &\quad+\frac12
 \sum_{\sigma,\tau\in\{\uparrow,\downarrow\}}
 \sum_{i,j,k,l}
 \left\langle e_i\otimes e_j,
 \frac{e_k\otimes e_l}{|x-y|}\right\rangle
 a_\sigma^*(e_i)a_\tau^*(e_j)
 a_\tau(e_l)a_\sigma(e_k),
 \label{eq:hamiltonian-fock}
\end{align}
where $h_Z=-\Delta/2-Z/|\cdot|$ is the one-particle Hamiltonian. We denote the first and second terms in \eqref{eq:hamiltonian-fock} by $\dG(h_Z)$ and $\mathbb W$, respectively. Thus $\mathbb H=\dG(h_Z)+\mathbb W$. We also define the number operators
$$
 \mathcal N_\uparrow
 :=\bigoplus_{n,m\geq0}n\1_{\mathcal K_{n,m}}
 =\sum_{j}a_\uparrow^*(e_j)a_\uparrow(e_j),\qquad
 \mathcal N_\downarrow
 :=\bigoplus_{n,m\geq0}m\1_{\mathcal K_{n,m}}
 =\sum_{j}a_\downarrow^*(e_j)a_\downarrow(e_j),
$$
and $\mathcal N:=\mathcal N_\uparrow+\mathcal N_\downarrow$.

\subsection{Spectrum of the hydrogenic Hamiltonian}

To choose an orthonormal basis $(e_j)_{j\geq0}$ of $\mathfrak h$, we recall the spectrum of the hydrogen Hamiltonian.

Equip $\mathbb S^2$ with surface measure of mass $4\pi$. The map $Uf(r,\omega)=rf(r\omega)$ is unitary from $L^2(\R^3)$ onto $L^2(\R_+,\dd r)\otimes L^2(\mathbb S^2)$. For the standard orthonormal spherical harmonics $Y_{\ell,m}$, let $\mathcal Y_\ell=\ker(-\Delta_{\mathbb S^2}-\ell(\ell+1))
=\operatorname{span}\{Y_{\ell,m}:|m|\leq\ell\}$. For $Z>0$, we have
\begin{equation}\label{eq:one-particle-angular-sum}
 L^2(\R^3)\cong\bigoplus_{\ell\geq0}
       (L^2(\R_+,\dd r)\otimes\mathcal Y_\ell),\qquad
 Uh_ZU^*=\bigoplus_{\ell\geq0}(h_{Z,\ell}\otimes \1_{\mathcal Y_\ell}).
\end{equation}
Here $h_{Z,\ell}$ is the Friedrichs realization of $-\tfrac12\d^2/\d r^2+\ell(\ell+1)/(2r^2)-Z/r$ with form domain $H_0^1(\R_+)$. The eigenvalue $\lambda_n(Z)=-Z^2/(2n^2)$ is simple for $n\geq\ell+1$. We choose real normalized eigenfunctions $R^Z_{n,\ell}$. For example, we may take
\begin{align*}
 R^Z_{1,0}(r)=2Z^{3/2}re^{-Zr},\quad
 R^Z_{2,0}(r)=\frac{Z^{3/2}}{2\sqrt2}r(2-Zr)e^{-Zr/2},\quad
 R^Z_{2,1}(r)=\frac{Z^{5/2}}{2\sqrt6}r^2e^{-Zr/2}.
\end{align*}
We recall the following spectral properties of the hydrogenic Hamiltonian $h_Z$~\cite{Teschl09}.

\begin{proposition}\label{lem:hydrogen}
For $Z>0$, the operator $h_Z=-\Delta/2-Z/|\cdot|$ is self-adjoint with form domain $H^1(\R^3)$. We have
\begin{align*}
 \sigma_{\mathrm{disc}}(h_Z)=\big\{\lambda_n(Z):n\geq1\big\},\qquad
 \sigma_{\mathrm{ess}}(h_Z)=[0,\infty),
\end{align*}
with eigenspace
\begin{equation}\label{eq:one-particle-eigenspaces}
 \Ran\big(\1_{\{\lambda_n(Z)\}}(h_Z)\big)=\operatorname{span}\big\{\psi^Z_{n,\ell,m}(r\omega)
     =r^{-1}R^Z_{n,\ell}(r)Y_{\ell,m}(\omega):
            0\leq\ell<n,\ |m|\leq\ell\big\}
\end{equation}
of dimension $n^2$.
\end{proposition}

For fixed $Z$, write $\lambda_n=\lambda_n(Z)$, $R_{n,\ell}=R_{n,\ell}^Z$ and $\psi_{n,\ell,m}=\psi_{n,\ell,m}^Z$. We fix an orthonormal basis $(e_j)_{j\geq0}$ of $\mathfrak h$ and we take all $\psi_{n,\ell,m}$ with $n\geq3$. For example,
$$
 e_0=\psi_{1,0,0},\qquad
 e_1=\psi_{2,0,0},\qquad
 e_2=\psi_{2,1,-1},\qquad
 e_3=\psi_{2,1,0},\qquad
 e_4=\psi_{2,1,1},\qquad\cdots
$$
In particular,
\begin{equation}\label{eq:def-of-e-0}
      e_0(x)=\sqrt{\frac{Z^3}{\pi}}\exp(-Z|x|).
\end{equation}
To complete the basis, we take an orthonormal basis of each subspace $\Ran\1_{[k,k+1)}(h_Z)$, $k=0,1,\ldots$.

\section{Proof strategy}\label{sec:strategy}

We first reduce the problem to $\mathcal K_N$ by a standard spin argument.

\begin{lemma}\label{lem:balanced}
For $N\geq1$, $Z>0$ and $b\in\R$,
$$
 H_{N,Z}\geq b\text{ on }\gH_N^{\F}
 \quad\Longleftrightarrow\quad
 H_{N,Z}^{\mathrm{sc}}\geq b\text{ on }\mathcal K_N:=\mathcal K_{\lceil N/2\rceil,\lfloor N/2\rfloor}.
$$
In particular,
$$
 E^{\F}(N,Z)
 =\inf\sigma\bigl(H_{N,Z}^{\mathrm{sc}}|_{\mathcal K_N}\bigr).
$$
If this energy is an eigenvalue, each of the sectors $\gH_{\lceil N/2\rceil,\lfloor N/2\rfloor}^{\F}$ and $\gH_{\lfloor N/2\rfloor,\lceil N/2\rceil}^{\F}$ contains a ground state.
\end{lemma}

\begin{proof}
Let
$$
 S_-=\sum_{j=1}^N
       \bigl(\ket{\downarrow}\bra{\uparrow}\bigr)_j,
 \qquad S_+=S_-^*,
$$
where $\{\uparrow,\downarrow\}$ is the standard basis of $\C^2$. The spin-sector projections, $S_\pm$, and simultaneous interchange of the spin labels commute with $H_{N,Z}$ and its spectral projections. Moreover, for $\psi\in\gH_{n,m}^{\F}$,
$$
 \|S_-\psi\|^2-\|S_+\psi\|^2
 =\ip\psi{[S_+,S_-]\psi}
 =(n-m)\|\psi\|^2.
$$
Thus $S_-\psi\neq0$ whenever $\psi\neq0$ and $n>m$. Take a nonzero spin-sector component in a nonzero spectral subspace. Interchange the spin labels if necessary and apply $S_-$ repeatedly until $n=\lceil N/2\rceil$. This gives a nonzero vector in the same spectral subspace and in $\gH_{\lceil N/2\rceil,\lfloor N/2\rfloor}^{\F}$. Applying this observation to the spectral interval $(-\infty,b)$ proves the equivalence and hence the equality of the spectral infima. Applying it to the ground eigenspace, followed by interchanging the spin labels, proves the last assertion.
\end{proof}

To prove Theorem~\ref{thm:main}, it suffices to establish the following estimate.

\begin{theorem}\label{thm:main-reduced}
For every $\psi\in H^1((\R^3)^3)\cap\mathcal K_{2,1}$, there exist $f\in H^1(\R^3)$ and $\eta\in H^1((\R^3)^3)\cap\mathcal K_{2,1}$, uniquely determined by $\psi$, such that
$$
 \ip\psi{\big(H_{3,2}^{\mathrm{sc}}-E^{\F}(2,2)\big)\psi}
 \geq\frac1{200}\norm{\nabla f}_2^2+\frac1{200}\norm\eta^2.
$$
Moreover, if $f=0$ in $H^1(\R^3)$ and $\eta=0$ in $H^1((\R^3)^3)\cap\mathcal K_{2,1}$, then $\psi=0$.
\end{theorem}
The proof is given in Section~\ref{sec:1-d-variational}.

\begin{proof}[Proof of Theorem~\ref{thm:main}]
Lemma~\ref{lem:balanced} and Theorem~\ref{thm:main-reduced} give
$$
 E^{\F}(3,2)=\inf\sigma\bigl(H_{N,Z}^{\mathrm{sc}}|_{\mathcal K_N}\bigr)\geq E^{\F}(2,2),
$$
while \eqref{eq:energy-monotonicity} gives the reverse inequality.

If a ground state existed in $\gH_3^{\F}$, Lemma~\ref{lem:balanced} would give a nonzero ground state $\psi$ in $\mathcal K_{2,1}$. The preceding estimate would give $\nabla f=0$ and $\eta=0$. Since $f\in L^2(\R^3)$, we would have $f=0$ and hence $\psi=0$, a contradiction.
\end{proof}

We now outline the proof of Theorem~\ref{thm:main-reduced}. A standard argument gives a unique normalized positive ground state $\varphi\in H^2(\R^6)$ for $H_{2,2}^{\mathrm{sc}}$ on $\mathcal K_{1,1}$. It is real, symmetric in its two variables, and invariant under simultaneous spatial rotations. Lemma~\ref{lem:balanced} gives
$$
 \ip{\phi}{H_{2,2}^{\mathrm{sc}}\phi}=\inf\sigma(H_{2,2}^{\mathrm{sc}})=E^{\F}(2,2).
$$

Our starting point is the usual shell picture of neutral helium, where two electrons with opposite spins occupy the same $s$ orbital. This suggests approximating $\varphi$ by $e_0\otimes e_0$. We measure the deviation from this product state by
\begin{equation}\label{eq:intro-overlap}
 1-\|a_\downarrow(e_0)a_\uparrow(e_0)\varphi\|^2
 =1-|\langle e_0\otimes e_0,\varphi\rangle|^2.
\end{equation}

Consider adding a third electron with spatial orbital $f$. We may assume that its spin is up. If the two-electron ground state were exactly $e_0\otimes e_0$, the Pauli exclusion principle would allow us to restrict to $f\perp e_0$, and the resulting unnormalized three-electron state would be $a_\uparrow^*(f)\varphi$. This motivates the decomposition
$$
 \psi=a_\uparrow^*(f)\varphi+\eta,
 \qquad f\perp e_0.
$$
By Proposition~\ref{prop:decomposition}, every $\psi\in H^1\cap\mathcal K_{2,1}$ admits this decomposition with $f\in H^1(\R^3)$ and $\eta\in H^1\cap\mathcal K_{2,1}$, uniquely determined by requiring $\eta$ to be orthogonal to all states $a_\uparrow^*(g)\varphi$ with $g\perp e_0$. The energy then splits as
\begin{align*}
 \ip\psi{(H_{3,2}^{\mathrm{sc}}-E^{\F}(2,2))\psi}
 &=
 \underbrace{
 \Bigl(
 \ip\eta{(H_{3,2}^{\mathrm{sc}}-E^{\F}(2,2))\eta}
 +2\operatorname{Re}
 \ip\eta{(H_{3,2}^{\mathrm{sc}}-E^{\F}(2,2))a_\uparrow^*(f)\varphi}
 \Bigr)
 }_{\mathrm{I}}\\
 &\quad+
 \underbrace{
 \ip{a_\uparrow^*(f)\varphi}
    {(H_{3,2}^{\mathrm{sc}}-E^{\F}(2,2))a_\uparrow^*(f)\varphi}
 }_{\mathrm{II}}.
\end{align*}

\subsection{Reduction to a one-particle quadratic form}

To bound $\mathrm{I}$, we reserve $\delta_\theta\|\eta\|^2$ from the remainder energy and complete the square with the remaining energy and the cross term. This gives
$$
 \mathrm{I}=
 \ip\eta{(H_{3,2}^{\mathrm{sc}}-E^{\F}(2,2))\eta}
 +2\operatorname{Re}
 \ip\eta{(H_{3,2}^{\mathrm{sc}}-E^{\F}(2,2))a_\uparrow^*(f)\varphi}
 \geq
 \delta_\theta\|\eta\|^2
 -\inf_{g_\uparrow+g_\downarrow\perp e_0}
       \mathcal C_{\Gamma_\theta}(f;g_\uparrow,g_\downarrow).
$$
Here $\theta$ is an adjustable parameter. The energy $E^{\F}(2,2)$, the spectral gap of $H_{2,2}^{\mathrm{sc}}$, and bounds on \eqref{eq:intro-overlap} control $\delta_\theta$. The nonnegative quadratic functional $\mathcal C_{\Gamma_\theta}$ measures the loss from controlling the cross term. See Sections~\ref{sec:core} and~\ref{sec:square} for the definitions and Propositions~\ref{prop:remainder} and~\ref{prop:completion-of-the-square} for the estimates.

We solve the constrained minimization explicitly. Part of the loss cancels terms in $\mathrm{II}$, and we retain the positive contribution from the minimization. The resulting one-particle quadratic form is
$$
 \mathrm{II}
 -\inf_{g_\uparrow+g_\downarrow\perp e_0}
       \mathcal C_{\Gamma_\theta}(f;g_\uparrow,g_\downarrow)
 =\mathfrak F_\theta[f].
$$
Thus,
$$
 \ip\psi{(H_{3,2}^{\mathrm{sc}}-E^{\F}(2,2))\psi}
 \geq \mathfrak F_\theta[f]+\delta_\theta\|\eta\|^2.
$$

\subsection{One-dimensional variational problem}

To bound $\mathfrak F_\theta[f]$ from below in terms of the kinetic energy of $f$, we use Proposition~\ref{prop:scalar-source-reduction}. It bounds the operator terms by explicit scalar potentials and gives
\begin{align*}
    \mathfrak F_\theta[f]
    \geq
    \frac12\|\nabla f\|_2^2
    -\tr{\Gamma_\theta}
    \int_{\mathbb R^3}
    \Bigl(
        \frac2{|x|}
        -\frac2{\sqrt{|x|^2+2}}
    \Bigr)|f(x)|^2\,dx
    -\frac{\tr\Gamma_\theta}{10}
 \int_{\R^3}
 \frac{|P_{\mathrm{odd}}f|^2}{(|x|^2+2)^{3/2}}\dd x.
\end{align*}
To absorb the negative potential terms into the kinetic energy under the constraint $f\perp e_0$, we define the optimal constant $C_*$ by
\begin{align*}
C_*:=\sup_{\substack{
f\perp e_0,\norm{\nabla f}_2^2=1
}}\Big\{\,\int_{\R^3}V|f|^2\dd x+\frac{1}{20}\int_{\R^3}\frac{|P_{\mathrm{odd}}f|^2}{(|x|^2+2)^{3/2}}\dd x\,\Big\},\qquad V(r):=\frac{1}{r}-\frac{1}{\sqrt{r^2+2}},
\end{align*}
which gives
$$
 \mathfrak F_\theta[f]
 \geq
 \frac{1-4C_*\tr\Gamma_\theta}{2}\|\nabla f\|_2^2.
$$
It suffices to bound $C_*$ from above.

By Lemma~\ref{lem:reduced-C-star}, we may restrict the supremum defining $C_*$ to radial functions. The substitution $f(x)=|x|^{-1}u(|x|)$ gives the one-dimensional variational problem
$$
 C_*=
 \sup_{\substack{
 0\ne u\in\dot H_0^1(0,\infty)\\
 \int_0^\infty re^{-2r}u(r)\dd r=0}}
 \frac{\displaystyle\int_0^\infty V(r)|u(r)|^2\dd r}
      {\displaystyle\int_0^\infty|u'(r)|^2\dd r},
 \qquad
 V(r)=\frac1r-\frac1{\sqrt{r^2+2}}.
$$
We write this constant as $M_V(2)$, with $M_W(\alpha)$ defined in \eqref{eq:def-of-M-A}.

We compare $V$ with a weight $W$ for which we can solve the constrained problem explicitly. Under the condition in Lemma~\ref{lem:log-convex-weight}, the second eigenfunction $v$ of the weighted problem determines a parameter $\alpha_W$ such that
$$
 \int_0^\infty re^{-\alpha_Wr}v(r)\dd r=0,
 \qquad
 M_W(\alpha_W)=\frac1{\lambda_2(W)}.
$$
We choose $W$ with an explicit second eigenfunction and rescale it to match the constraint. The scaling relation \eqref{eq:rescale} gives
$$
 C_*=M_V(2)
 \leq M_W(2b)\sup_{r>0}\frac{V(r)}{W_b(r)},
 \qquad
 W_b(r)=\frac1{b^2}W\Bigl(\frac rb\Bigr).
$$
Taking $b=\alpha_W/2$ makes the first factor explicit and reduces the estimate to a pointwise comparison of the weights.

\section{Orthogonal decomposition}\label{sec:decomposition}

\begingroup Empty operator products are understood as the identity.

\begin{proposition}\label{prop:decomposition}
For $n\geq 1$ and $m\geq 0$, let $\varphi\in\mathcal K_{n-1,m}$ be normalized. Choose $v_1,\ldots,v_{n+m-1}\in\mathfrak h$ that are orthonormal within each group $v_1,\ldots,v_{n-1}$ and $v_n,\ldots,v_{n+m-1}$ and satisfy
$$
 \big\|a_\downarrow(v_{n+m-1})\cdots a_\downarrow(v_n)
 a_\uparrow(v_{n-1})\cdots a_\uparrow(v_1)\varphi\big\|>0.
$$
Then every $\psi\in\mathcal K_{n,m}$ has a unique orthogonal decomposition
\begin{equation}\label{eq:general-decomposition}
 \begin{gathered}
 \psi=a_\uparrow^*(f)\varphi+\eta,\qquad
 f\perp v_1,\ldots,v_{n-1},\\
 \ip\varphi{a_\uparrow(g)\eta}=0
 \qquad(\forall g\perp v_1,\ldots,v_{n-1}),\\
 \ip\varphi{a_\downarrow(g)\eta}=0
 \qquad(\forall g\in\mathfrak h).
 \end{gathered}
\end{equation}
Here the pairings are taken in $\mathcal F_{\mathrm f}$. The remainder $\eta$ satisfies
\begin{align}\label{eq:general-occupation}
 &\big\|a_\downarrow(v_{n+m-1})\cdots a_\downarrow(v_n)
 a_\uparrow(v_{n-1})\cdots a_\uparrow(v_1)\eta\big\|^2\notag\\
 &\quad\leq
 \left(1-\big\|a_\downarrow(v_{n+m-1})\cdots a_\downarrow(v_n)
 a_\uparrow(v_{n-1})\cdots a_\uparrow(v_1)\varphi\big\|^2\right)
 \|\eta\|^2.
\end{align}
\end{proposition}

\begin{proof}
For $f\perp v_1,\ldots,v_{n-1}$, the CAR and the norm bounds give
\begin{align*}
 \|a_\uparrow^*(f)\varphi\|^2
 &\geq\big\|a_\downarrow(v_{n+m-1})\cdots a_\downarrow(v_n)
 a_\uparrow(v_{n-1})\cdots a_\uparrow(v_1)a_\uparrow^*(f)\varphi\big\|^2\\
 &=\big\|a_\downarrow(v_{n+m-1})\cdots a_\downarrow(v_n)
 a_\uparrow(v_{n-1})\cdots a_\uparrow(v_1)\varphi\big\|^2\|f\|^2.
\end{align*}
Here the contraction of $\varphi$ is a scalar. Thus $f\mapsto a_\uparrow^*(f)\varphi$ is bounded below on $\{v_1,\ldots,v_{n-1}\}^\perp\subset\mathfrak h$ and has closed range. Orthogonal projection onto this range gives the unique decomposition with the stated condition on $a_\uparrow(g)\eta$. For every $g\in\mathfrak h$,
\begin{align*}
 \ip\varphi{a_\downarrow(g)\eta}
 &=\ip{a_\downarrow^*(g)\varphi}\eta
 =0,
\end{align*}
since $a_\downarrow^*(g)\varphi\in\mathcal K_{n-1,m+1}
\perp\mathcal K_{n,m}\ni\eta$. This proves \eqref{eq:general-decomposition}.

For $g\perp v_1,\ldots,v_{n-1}$, we have $a_\uparrow(g)\eta\perp\varphi$. Applying Cauchy--Schwarz yields
\begin{align*}
 &\left|\ip{g}{a_\downarrow(v_{n+m-1})\cdots a_\downarrow(v_n)
 a_\uparrow(v_{n-1})\cdots a_\uparrow(v_1)\eta}\right|^2\\
 &\quad=
\left|
\ip{\bigl(\1-|\varphi\rangle\langle\varphi|\bigr)
a_\uparrow^*(v_1)\cdots a_\uparrow^*(v_{n-1})
a_\downarrow^*(v_n)\cdots a_\downarrow^*(v_{n+m-1})1}{a_\uparrow(g)\eta}
\right|^2\\
&\quad\leq
\left\|
\bigl(\1-|\varphi\rangle\langle\varphi|\bigr)
a_\uparrow^*(v_1)\cdots a_\uparrow^*(v_{n-1})
a_\downarrow^*(v_n)\cdots a_\downarrow^*(v_{n+m-1})1
\right\|^2
\|a_\uparrow(g)\eta\|^2\\
&\quad=
\left(
1-
\left\|
a_\downarrow(v_{n+m-1})\cdots a_\downarrow(v_n)
a_\uparrow(v_{n-1})\cdots a_\uparrow(v_1)\varphi
\right\|^2
\right)
\|a_\uparrow(g)\eta\|^2\\
&\quad\leq
\left(
1-
\left\|
a_\downarrow(v_{n+m-1})\cdots a_\downarrow(v_n)
a_\uparrow(v_{n-1})\cdots a_\uparrow(v_1)\varphi
\right\|^2
\right)
\|g\|^2\|\eta\|^2.
\end{align*}
The CAR imply that the vector $a_\downarrow(v_{n+m-1})\cdots a_\downarrow(v_n)
a_\uparrow(v_{n-1})\cdots a_\uparrow(v_1)\eta$ lies in $\mathcal K_{1,0}=\mathfrak h$ and is orthogonal to $v_1,\ldots,v_{n-1}$. Taking the supremum over unit vectors $g\perp v_1,\ldots,v_{n-1}$ proves \eqref{eq:general-occupation}.
\end{proof}
\endgroup

\begin{lemma}\label{lem:field-contraction-count}
Let $B$ be a normally ordered polynomial with $r$ creation operators and $s$ annihilation operators in each monomial. Then
\begin{align}
 \sum_{\sigma,j}a_{\sigma}^*(e_j)Ba_{\sigma}(e_j)
 =(-1)^{r+s}B(\mathcal N-s)
 =(-1)^{r+s}(\mathcal N-r)B.
 \label{eq:field-contraction-count}
\end{align}
If the numbers $r_\sigma,s_\sigma$ of creation and annihilation operators of spin $\sigma$ are also the same in every monomial, then
\begin{align}
 \sum_{j}a_{\sigma}^*(e_j)Ba_{\sigma}(e_j)
 =(-1)^{r+s}B(\mathcal N_\sigma-s_\sigma)
 =(-1)^{r+s}(\mathcal N_\sigma-r_\sigma)B.
 \label{eq:field-contraction-spin-count}
\end{align}
\end{lemma}

\begin{proof}
 It suffices to consider a single monomial. Both identities follow from the CAR.
\end{proof}

\section{Remainder estimate}\label{sec:core}

From now on, assume $Z>5/4$. On $\mathcal K_{1,1}=L^2(\R^6)$, the Coulomb operator $H_{2,Z}^{\mathrm{sc}}$ is self-adjoint on $H^2(\R^6)$, with $C_c^\infty(\R^6)$ as an operator core \cite[Theorem~11.1]{Teschl09}. It is bounded below, and the HVZ theorem gives $\inf\sigma_{\mathrm{ess}}(H_{2,Z}^{\mathrm{sc}})=-Z^2/2$. Lemma~\ref{lem:balanced} and the normalized trial function $e_0\otimes e_0$, with $e_0$ as in \eqref{eq:def-of-e-0}, give
$$
 E^{\F}(2,Z)=\inf\sigma(H_{2,Z}^{\mathrm{sc}})
 \leq\langle e_0\otimes e_0,
             H_{2,Z}^{\mathrm{sc}}(e_0\otimes e_0)\rangle
 =-Z^2+\frac{5Z}{8}<-\frac{Z^2}{2}=\lambda_1.
$$
Thus $E^{\F}(2,Z)$ is an isolated eigenvalue. By the positivity theorem \cite[Theorem~10.12]{Teschl09}, it is simple and has a unique normalized positive eigenfunction $\varphi\in H^2(\R^6)$. Exchange of the two variables and simultaneous spatial rotations commute with $H_{2,Z}^{\mathrm{sc}}$ and preserve positivity and normalization. Uniqueness therefore gives $\varphi(y,x)=\varphi(x,y)$ and $\varphi(Ox,Oy)=\varphi(x,y)$ for every $O\in\mathrm{SO}(3)$.

Let $\lambda_1<\lambda_2$ be the first two eigenvalues of $h_Z$, with $h_Ze_0=\lambda_1e_0$. We use parameters $\mu\in\R$ and $0\leq\varepsilon_0<1$ satisfying
\begin{equation}\label{eq:core-inputs-main}
 H_{2,Z}^{\mathrm{sc}}\big|_{\varphi^\perp}\geq\mu,\qquad
 \norm{a_\downarrow(e_0)a_\uparrow(e_0)\varphi}^2
 \geq1-\varepsilon_0.
\end{equation}
Define the integral operator and its one-particle density by
\begin{equation}\label{eq:core-density}
 (\mathcal R_\varphi f)(x)=\int_{\R^3}\varphi(x,y)f(y)\dd y,
 \qquad \gamma_\varphi=\mathcal R_\varphi^2,
 \qquad \tr\gamma_\varphi=1.
\end{equation}
Thus $\mathcal R_\varphi$ is a real self-adjoint Hilbert--Schmidt operator, and $\gamma_\varphi$ is positive. We also set $P_0=\ket{e_0}\bra{e_0}$ and $Q_0=\1-P_0$, and define
$$
 \mathcal N_0=\sum_{\sigma\in\{\uparrow,\downarrow\}}
                  a_\sigma^*(e_0)a_\sigma(e_0).
$$

Proposition~\ref{prop:decomposition}, with $n=2$, $m=1$ and $v_1=v_2=e_0$, gives the unique orthogonal decomposition
\begin{equation}\label{eq:remainder-conditions}
 \begin{gathered}
 \psi=a_\uparrow^*(f)\varphi+\eta,\qquad f\perp e_0,\\
 \ip\varphi{a_\sigma(g)\eta}=0\qquad(\sigma\in\{\uparrow,\downarrow\},\ g\perp e_0),
 \end{gathered}
\end{equation}
for every $\psi\in\mathcal K_{2,1}$. If $\psi\in H^1$, then $f\in H^1(\R^3)$ and $\eta\in H^1\cap\mathcal K_{2,1}$.

\begin{proposition}\label{prop:remainder}
Assume that $\lambda_1+\lambda_2\leq\mu<2\lambda_2$. For $0<\theta\leq\min\{\frac{\lambda_2-\lambda_1}{2\lambda_2-\lambda_1-\mu/2},1\}$, define
\begin{align}
 \delta_\theta
 :=(1-\theta)\lambda_1+(2-\theta)\lambda_2+\theta\mu-E^{\F}(2,Z)+\varepsilon_0
 \left(\lambda_1-\lambda_2
 +\theta\left(\lambda_2-\frac\mu2\right)\right).
 \label{eq:remainder-coefficient}
\end{align}
Let $n+m=3$ and let $\eta\in H^1\cap\mathcal K_{n,m}$ satisfy \eqref{eq:remainder-conditions}. Then
\begin{align}
 \ip\eta{(H_{3,Z}^{\mathrm{sc}}-E^{\F}(2,Z))\eta}
 \geq\Big(1-\frac\theta2\Big)\sum_{\sigma,j\neq0}
 \ip{a_\sigma(e_j)\eta}{\mathbb W a_\sigma(e_j)\eta}+\sum_\sigma
 \ip{a_\sigma(e_0)\eta}{\mathbb W a_\sigma(e_0)\eta}
 +\delta_\theta\norm\eta^2.
 \label{eq:remainder-bound}
\end{align}
\end{proposition}

\begin{proof}
Lemma~\ref{lem:field-contraction-count} yields
\begin{align}
 \sum_{\sigma,r}a_\sigma^*(e_r)\dG(h_Z)a_\sigma(e_r)
 =(\mathcal N-1)\dG(h_Z),\qquad
 \sum_{\sigma,r}a_\sigma^*(e_r)\mathbb W a_\sigma(e_r)
 =(\mathcal N-2)\mathbb W.
 \label{eq:annihilation-one-and-two-body}
\end{align}
Since $h_Ze_0=\lambda_1e_0$, we also have $[a_\sigma(e_0),\dG(h_Z)]=a_\sigma(h_Ze_0)=\lambda_1a_\sigma(e_0)$. Thus
\begin{align*}
 &\sum_\sigma a_\sigma^*(e_0)\dG(h_Z)a_\sigma(e_0)
 =\sum_\sigma a_\sigma^*(e_0)
       \bigl(a_\sigma(e_0)\dG(h_Z)
                              -\lambda_1a_\sigma(e_0)\bigr)
                              =\mathcal N_0\dG(h_Z)-\lambda_1\mathcal N_0.
\end{align*}
Subtracting the $j=0$ terms from the full sums yields
\begin{align}
 \sum_{\sigma,j\neq0}a_\sigma^*(e_j)\mathbb H a_\sigma(e_j)
 &=\sum_{\sigma,j}a_\sigma^*(e_j)
                    \bigl(\dG(h_Z)+\mathbb W\bigr)a_\sigma(e_j)
       -\sum_\sigma a_\sigma^*(e_0)
                    \bigl(\dG(h_Z)+\mathbb W\bigr)a_\sigma(e_0)\nn\\
 &=(\mathcal N-1-\mathcal N_0)
                    \dG(h_Z)+\lambda_1\mathcal N_0+(\mathcal N-2)\mathbb W
       -\sum_\sigma a_\sigma^*(e_0)\mathbb W a_\sigma(e_0).
 \label{eq:annihilation-energy}
\end{align}
Combining \eqref{eq:annihilation-one-and-two-body} and \eqref{eq:annihilation-energy} gives the identity on $\mathcal F_{\mathrm f}$
\begin{align}
 &\mathbb H
 -\Bigl(1-\frac\theta2\Bigr)
     \sum_{\sigma,j\neq0}a_\sigma^*(e_j)\mathbb W a_\sigma(e_j)
 -\sum_\sigma a_\sigma^*(e_0)\mathbb W a_\sigma(e_0)\nn\\
 &\quad=\mathbb H
 -\Bigl(1-\frac\theta2\Bigr)
     \sum_{\sigma,j}a_\sigma^*(e_j)\mathbb W a_\sigma(e_j)
 -\frac\theta2\sum_\sigma a_\sigma^*(e_0)\mathbb W a_\sigma(e_0)\nn\\
 &\quad=\dG(h_Z)
       +(3-\mathcal N)\mathbb W
       +\frac\theta2\Bigl(
       (\mathcal N-2)\mathbb W
       -\sum_\sigma a_\sigma^*(e_0)\mathbb W a_\sigma(e_0)\Bigr)\nn\\
 &\quad=\Bigl(1-\frac\theta2
       (\mathcal N-1-\mathcal N_0)\Bigr)
       \dG(h_Z)-\frac{\theta\lambda_1}{2}\mathcal N_0
       +\frac\theta2\sum_{\sigma,j\neq0}
                  a_\sigma^*(e_j)\mathbb H a_\sigma(e_j)
       +(3-\mathcal N)\mathbb W.
 \label{eq:remainder-interpolation}
\end{align}
For $\eta\in\mathcal K_{n,m}$ with $n+m=3$, we have $\mathcal N\eta=3\eta$ and $\ip\eta{(3-\mathcal N) \mathbb W\eta}=0$. For $j\neq0$, \eqref{eq:remainder-conditions} gives $\ip\varphi{a_\sigma(e_j)\eta}=0$ whenever the contraction belongs to $\mathcal K_{1,1}$. Thus
\begin{align*}
 \sum_{\sigma,j\neq0}
       \ip{a_\sigma(e_j)\eta}{\mathbb H a_\sigma(e_j)\eta}
 \geq\mu\sum_{\sigma,j\neq0}\norm{a_\sigma(e_j)\eta}^2
 =\mu\ip\eta{
       (\mathcal N-\mathcal N_0)\eta}
 =\mu\ip\eta{(3-\mathcal N_0)\eta}.
\end{align*}
Combining this bound with \eqref{eq:remainder-interpolation} gives
\begin{align}
 &\ip\eta{(H_{3,Z}^{\mathrm{sc}}-E^{\F}(2,Z))\eta}
 -\Bigl(1-\frac\theta2\Bigr)\sum_{\sigma,j\neq0}
       \ip{a_\sigma(e_j)\eta}{\mathbb W a_\sigma(e_j)\eta}
 -\sum_\sigma
       \ip{a_\sigma(e_0)\eta}{\mathbb W a_\sigma(e_0)\eta}\nn\\
 &\quad=\ip\eta{\Bigl(
       \Bigl(1-\theta+\frac\theta2\mathcal N_0\Bigr)\dG(h_Z)
       -\frac{\theta\lambda_1}{2}\mathcal N_0-E^{\F}(2,Z)\Bigr)\eta}
       +\frac\theta2\sum_{\sigma,j\neq0}
       \ip{a_\sigma(e_j)\eta}{\mathbb H a_\sigma(e_j)\eta}\nn\\
 &\quad\geq\ip\eta{\Bigl(
       \Bigl(1-\theta+\frac\theta2\mathcal N_0\Bigr)\dG(h_Z)
       -\frac{\theta\lambda_1}{2}\mathcal N_0
       +\frac{\theta\mu}{2}(3-\mathcal N_0)-E^{\F}(2,Z)\Bigr)\eta}.\label{eq:remainder-ip-with-eta}
\end{align}

The spectral decomposition $h_Z=\lambda_2 \1+(\lambda_1-\lambda_2)\ket{e_0}\bra{e_0}+(h_Z-\lambda_2)_+$ gives the Fock-space identity
\begin{align*}
 \dG(h_Z)
 =\lambda_2\mathcal N+(\lambda_1-\lambda_2)\mathcal N_0+\dG((h_Z-\lambda_2)_+).
\end{align*}
Since $h_Ze_0=\lambda_1e_0$, we have $[\mathcal N_0,\dG(h_Z)]=[\mathcal N_0,\dG((h_Z-\lambda_2)_+)]=0$. Together with $1-\theta+\theta\mathcal N_0/2\geq1-\theta\geq0$, this gives
\begin{align}
 \Bigl(1-\theta+\frac\theta2\mathcal N_0\Bigr)
 \dG((h_Z-\lambda_2)_+)
 =\Bigl(1-\theta+\frac\theta2\mathcal N_0\Bigr)^{1/2}
 \dG((h_Z-\lambda_2)_+)
 \Bigl(1-\theta+\frac\theta2\mathcal N_0\Bigr)^{1/2}
 \geq0.\label{eq:h-2lambda-2-positive}
\end{align}
We expand the resulting polynomial in $\mathcal N_0$ to obtain
\begin{align}
 &\Bigl(1-\theta+\frac\theta2\mathcal N_0\Bigr)
       \bigl(3\lambda_2+(\lambda_1-\lambda_2)\mathcal N_0\bigr)
       -\frac{\theta\lambda_1}{2}\mathcal N_0
       +\frac{\theta\mu}{2}(3-\mathcal N_0)\nn\\
 &\quad=\Bigl(3(1-\theta)\lambda_2+\frac{3\theta\mu}{2}\Bigr)\1
       +\Bigl((1-\theta)\lambda_1-(1-2\theta)\lambda_2
                           -\frac{\theta\mu}{2}\Bigr)\mathcal N_0+\frac{\theta(\lambda_1-\lambda_2)}{2}(\mathcal N_0-1)\mathcal N_0\nn\\
 &\quad=\Bigl((1-\theta)\lambda_1+(2-\theta)\lambda_2
                     +\theta\mu\Bigr)\1+\Bigl(\lambda_1-\lambda_2
                     +\theta\Bigl(\lambda_2-\frac\mu2\Bigr)\Bigr)
                    \frac{(\mathcal N_0-1)\mathcal N_0}2\nn\\
 &\qquad+\Bigl(\lambda_2-\lambda_1
              -\theta\Bigl(2\lambda_2-\lambda_1-\frac\mu2\Bigr)\Bigr)
                    \frac{(\mathcal N_0-1)(\mathcal N_0-2)}2\nn\\
 &\quad\geq\Bigl((1-\theta)\lambda_1+(2-\theta)\lambda_2
                     +\theta\mu\Bigr)\1+\Bigl(\lambda_1-\lambda_2
                     +\theta\Bigl(\lambda_2-\frac\mu2\Bigr)\Bigr)
                    \frac{(\mathcal N_0-1)\mathcal N_0}2,
 \label{eq:remainder-occupation}
\end{align}
where the last inequality uses $\lambda_2-\lambda_1-\theta\Bigl(2\lambda_2-\lambda_1-\frac\mu2\Bigr)\geq 0$ and $\sigma(\mathcal N_0)=\{0,1,2\}$. Combining \eqref{eq:remainder-occupation} and \eqref{eq:h-2lambda-2-positive} with \eqref{eq:remainder-ip-with-eta} gives
\begin{align*}
 &\ip\eta{(H_{3,Z}^{\mathrm{sc}}-E^{\F}(2,Z))\eta}
 -\Bigl(1-\frac\theta2\Bigr)\sum_{\sigma,j\neq0}
       \ip{a_\sigma(e_j)\eta}{\mathbb W a_\sigma(e_j)\eta}
 -\sum_\sigma
       \ip{a_\sigma(e_0)\eta}{\mathbb W a_\sigma(e_0)\eta}\\
 &\quad\geq\ip\eta{\Bigl(
       \Bigl(1-\theta+\frac\theta2\mathcal N_0\Bigr)
          \bigl(3\lambda_2+(\lambda_1-\lambda_2)\mathcal N_0\bigr)
       -\frac{\theta\lambda_1}{2}\mathcal N_0
       +\frac{\theta\mu}{2}(3-\mathcal N_0)-E^{\F}(2,Z)\Bigr)\eta}\\
 &\quad\geq\Bigl((1-\theta)\lambda_1+(2-\theta)\lambda_2
                              +\theta\mu-E^{\F}(2,Z)\Bigr)\norm\eta^2
                              +\Bigl(\lambda_1-\lambda_2
                     +\theta\Bigl(\lambda_2-\frac\mu2\Bigr)\Bigr)
                    \ip\eta{\frac{(\mathcal N_0-1)\mathcal N_0}2\eta}\\
 &\quad=\Bigl((1-\theta)\lambda_1+(2-\theta)\lambda_2
                              +\theta\mu-E^{\F}(2,Z)\Bigr)\norm\eta^2+\Bigl(\lambda_1-\lambda_2
                     +\theta\Bigl(\lambda_2-\frac\mu2\Bigr)\Bigr)
                    \norm{a_\downarrow(e_0)a_\uparrow(e_0)\eta}^2\\
 &\quad\geq\Bigl((1-\theta)\lambda_1+(2-\theta)\lambda_2
                    +\theta\mu-E^{\F}(2,Z)
       +\varepsilon_0\Bigl(\lambda_1-\lambda_2
                     +\theta\Bigl(\lambda_2-\frac\mu2\Bigr)\Bigr)\Bigr)
                    \norm\eta^2=\delta_\theta\norm\eta^2.
\end{align*}
\end{proof}

\section{Completion of the square and the one-particle quadratic form}\label{sec:square}

Throughout this section, $\delta_\theta$ is as in \eqref{eq:remainder-coefficient}. For a positive one-particle density $\rho$, define the multiplication operators $D_\rho,A_\rho$ and the integral operators $K_\rho,L_\rho$ by
\begin{align}
 D_\rho(x)&=\int_{\R^3}\frac{\rho(y,y)}{|x-y|}\dd y,
 &K_\rho(x,y)&=\frac{\rho(x,y)}{|x-y|},\nn\\
 A_\rho(x)&=\int_{\R^3}\rho(y,y)|x-y|\dd y,
 &L_\rho(x,y)&=\rho(x,y)|x-y|.
 \label{eq:density-operators}
\end{align}
We define the cost functional
\begin{align}
 \mathcal C_\rho(f;g_\uparrow,g_\downarrow)
 &:={}\ip f{(2D_\rho-K_\rho)f}
 -2\tr\rho\,\operatorname{Re}\ip{g_\uparrow+g_\downarrow}f\nn\\
 &\quad+2\operatorname{Re}\ip{g_\downarrow}{\rho f}
 +\ip{g_\uparrow}{A_\rho g_\uparrow}
 +\ip{g_\downarrow}{(A_\rho-L_\rho)g_\downarrow},
 \label{eq:source-cost}
\end{align}
for $f,g_\uparrow,g_\downarrow\in\mathfrak h$. The functional is linear in $\rho$. For $\rho=\sum_j\ket{w_j}\bra{w_j}$, the CAR show that the functional is nonnegative through the representation
\begin{align}
 \mathcal C_\rho(f;g_\uparrow,g_\downarrow)
 &=\ip{f}{D_\rho f}-2\tr\rho\,\operatorname{Re}\ip{g_\uparrow}{f}+\ip{g_\uparrow}{A_\rho g_\uparrow}\nn\\
 &\quad+\ip{f}{(D_\rho-K_\rho) f}-2\tr\rho\,\operatorname{Re}\ip{g_\downarrow}{f}+2\operatorname{Re}\ip{g_\downarrow}{\rho f}+\ip{g_\downarrow}{(A_\rho-L_\rho)g_\downarrow}\nn\\
 &=\sum_j\norm{
 \mathbb W^{1/2}a_\uparrow^*(f)a_\downarrow^*(w_j)1
 -\mathbb W^{-1/2}a_\uparrow^*(g_\uparrow)
                         a_\downarrow^*(w_j)1}^2\nn\\
 &\quad+\sum_j\norm{
 \mathbb W^{1/2}a_\uparrow^*(f)a_\uparrow^*(w_j)1
 -\mathbb W^{-1/2}a_\uparrow^*(g_\downarrow)
                         a_\uparrow^*(w_j)1}^2.
 \label{eq:source-norms}
\end{align}
Here $\mathbb W$ acts in the two-particle sector. We also define
\begin{equation}\label{eq:source-density}
 \Gamma_{0,\phi}=\mathcal R_\varphi P_0\mathcal R_\varphi,\qquad
 \Gamma_{+,\phi}=\mathcal R_\varphi(\1-P_0)\mathcal R_\varphi,\qquad
 \Gamma_{\theta}=\Gamma_{0,\phi}+\frac{2}{2-\theta}\Gamma_{+,\phi}=\gamma_\phi+\frac{\theta}{2-\theta}\Gamma_{+,\phi}.
\end{equation}

\begin{proposition}[Completion of the square]
\label{prop:completion-of-the-square}
Let $\psi\in H^1\cap\mathcal K_{2,1}$ have the decomposition \eqref{eq:remainder-conditions}. Then for every $g_\uparrow,g_\downarrow\in\mathfrak h$ with $g_\uparrow+g_\downarrow\perp e_0$,
\begin{equation}
      \ip\eta{(\mathbb H-E^{\F}(2,Z))\eta}+2\operatorname{Re}\ip\eta{(\mathbb H-E^{\F}(2,Z))a_\uparrow^*(f)\varphi}\geq\delta_\theta\norm\eta^2-\mathcal C_{\Gamma_\theta}(f;g_\uparrow,g_\downarrow),
\end{equation}
where $\delta_\theta$ is given by \eqref{eq:remainder-coefficient}.
\end{proposition}

\begin{proof}
Lemma~\ref{lem:field-contraction-count} gives
\begin{equation}
 \sum_{\sigma,j}a_\sigma^*(e_j)
       [\mathbb W,a_\uparrow^*(f)]a_\sigma(e_j)
 =-[\mathbb W,a_\uparrow^*(f)](\mathcal N-1),\quad
 \sum_j a_\sigma^*(e_j)a_\uparrow^*(g_\sigma)a_\sigma(e_j)
 =-a_\uparrow^*(g_\sigma)\mathcal N_\sigma.
\end{equation}
Since $\mathcal N_\uparrow\varphi
=\mathcal N_\downarrow\varphi=\varphi$ and $\mathbb W a_\sigma(e_j)\varphi=0$, we obtain
\begin{align}
 \bigl([\mathbb W,a_\uparrow^*(f)]
             -a_\uparrow^*(g_\uparrow+g_\downarrow)\bigr)\varphi
 &=-\sum_{\sigma,j}a_\sigma^*(e_j)
                [\mathbb W,a_\uparrow^*(f)]a_\sigma(e_j)\varphi
       +\sum_{\sigma,j}a_\sigma^*(e_j)
                a_\uparrow^*(g_\sigma)a_\sigma(e_j)\varphi\nn\\
 &=-\sum_{\sigma,j}a_\sigma^*(e_j)
       \bigl([\mathbb W,a_\uparrow^*(f)]-a_\uparrow^*(g_\sigma)\bigr)
                                      a_\sigma(e_j)\varphi\nn\\
 &=-\sum_{\sigma,j}a_\sigma^*(e_j)
       \bigl(\mathbb W a_\uparrow^*(f)-a_\uparrow^*(g_\sigma)\bigr)
                                      a_\sigma(e_j)\varphi.
 \label{eq:source-factorization}
\end{align}
The commutator $[\dG(h_Z),a_\sigma^*(f)]=a_\sigma^*(h_Zf)$ and the orthogonality $a_\uparrow^*(g_\uparrow+g_\downarrow)\varphi\perp\eta$ give
\begin{align}
 \ip\eta{(\mathbb H-E^{\F}(2,Z))a_\uparrow^*(f)\varphi}
 &=\ip\eta{[\mathbb H,a_\uparrow^*(f)]\varphi}
 =\ip\eta{[\mathbb W,a_\uparrow^*(f)]\varphi}\nn\\
 &=\ip\eta{\bigl([\mathbb W,a_\uparrow^*(f)]
               -a_\uparrow^*(g_\uparrow+g_\downarrow)\bigr)\varphi}\nn\\
 &=-\sum_{\sigma,j}\ip{a_\sigma(e_j)\eta}
       {\bigl(\mathbb W a_\uparrow^*(f)-a_\uparrow^*(g_\sigma)\bigr)
                                      a_\sigma(e_j)\varphi}\nn\\
 &=-\sum_{\sigma,j}\left\langle
       \mathbb W^{1/2}a_\sigma(e_j)\eta,
       \mathbb W^{-1/2}
       \bigl(\mathbb W a_\uparrow^*(f)-a_\uparrow^*(g_\sigma)\bigr)
                                      a_\sigma(e_j)\varphi
                         \right\rangle.
 \label{eq:completion-mixed-pairing}
\end{align}

Using \eqref{eq:completion-mixed-pairing} and Proposition~\ref{prop:remainder}, we complete the square in each group to obtain
\begin{align}
 &\ip\eta{(\mathbb H-E^{\F}(2,Z))\eta}
      +2\operatorname{Re}\ip\eta{(\mathbb H-E^{\F}(2,Z))a_\uparrow^*(f)\varphi}
 \nn\\
 &\quad\geq\delta_\theta\norm\eta^2
       +\sum_\sigma\norm{\mathbb W^{1/2}a_\sigma(e_0)\eta}^2
       +\Bigl(1-\frac\theta2\Bigr)
              \sum_{\sigma,j\neq0}\norm{
                          \mathbb W^{1/2}a_\sigma(e_j)\eta}^2\nn\\
 &\qquad-2\operatorname{Re}\sum_{\sigma,j}\left\langle
       \mathbb W^{1/2}a_\sigma(e_j)\eta,
       \mathbb W^{-1/2}
       \bigl(\mathbb W a_\uparrow^*(f)-a_\uparrow^*(g_\sigma)\bigr)
                                      a_\sigma(e_j)\varphi
                         \right\rangle\nn\\
 &\quad=\delta_\theta\norm\eta^2
       +\sum_\sigma\norm{
       \mathbb W^{1/2}a_\sigma(e_0)\eta
       -\mathbb W^{-1/2}
       \bigl(\mathbb W a_\uparrow^*(f)-a_\uparrow^*(g_\sigma)\bigr)
                                      a_\sigma(e_0)\varphi}^2\nn\\
 &\qquad+\Bigl(1-\frac\theta2\Bigr)
          \sum_{\sigma,j\neq0}\norm{
       \mathbb W^{1/2}a_\sigma(e_j)\eta
       -\frac2{2-\theta}\mathbb W^{-1/2}
       \bigl(\mathbb W a_\uparrow^*(f)-a_\uparrow^*(g_\sigma)\bigr)
                                      a_\sigma(e_j)\varphi}^2\nn\\
 &\qquad-\sum_\sigma\norm{
       \mathbb W^{-1/2}
       \bigl(\mathbb W a_\uparrow^*(f)-a_\uparrow^*(g_\sigma)\bigr)
                                      a_\sigma(e_0)\varphi}^2\nn\\
 &\qquad-\frac2{2-\theta}\sum_{\sigma,j\neq0}\norm{
       \mathbb W^{-1/2}
       \bigl(\mathbb W a_\uparrow^*(f)-a_\uparrow^*(g_\sigma)\bigr)
                                      a_\sigma(e_j)\varphi}^2\nn\\
 &\quad\geq\delta_\theta\norm\eta^2-\sum_\sigma\norm{
       \mathbb W^{-1/2}
       \bigl(\mathbb W a_\uparrow^*(f)-a_\uparrow^*(g_\sigma)\bigr)
                                      a_\sigma(e_0)\varphi}^2\nn\\
 &\qquad-\frac2{2-\theta}\sum_{\sigma,j\neq0}\norm{
       \mathbb W^{-1/2}
       \bigl(\mathbb W a_\uparrow^*(f)-a_\uparrow^*(g_\sigma)\bigr)
                                      a_\sigma(e_j)\varphi}^2.
 \label{eq:weighted-pair-squares}
\end{align}
By \eqref{eq:source-norms} and \eqref{eq:source-density}, the last two lines of \eqref{eq:weighted-pair-squares} equal $\delta_\theta\norm{\eta}^2-\mathcal C_{\Gamma_\theta}(f;g_\uparrow,g_\downarrow)$.
\end{proof}

\begin{proposition}
\label{prop:symmetrized-square}
Let $\psi\in H^1\cap\mathcal K_{2,1}$ have the decomposition \eqref{eq:remainder-conditions}. Then
\begin{align}
 \ip{a_\uparrow^*(f)\varphi}
       {(H_{3,Z}^{\mathrm{sc}}-E^{\F}(2,Z))a_\uparrow^*(f)\varphi}
 =\ip f{(h_Z+2D_{\gamma_\varphi}-K_{\gamma_\varphi})f}
 +\ip{\mathcal R_\varphi f}{(h_Z-E^{\F}(2,Z))\mathcal R_\varphi f}.
 \label{eq:exact-hole}
\end{align}
Moreover,
\begin{align}
 \ip\psi{(H_{3,Z}^{\mathrm{sc}}-E^{\F}(2,Z))\psi}
 &\geq\ip f{(h_Z+2D_{\gamma_\varphi}-K_{\gamma_\varphi})f}
 +\ip{\mathcal R_\varphi f}{(h_Z-E^{\F}(2,Z))\mathcal R_\varphi f}\nn\\
 &\quad-\inf_{g_\uparrow+g_\downarrow\perp e_0}\mathcal C_{\Gamma_\theta}(f;g_\uparrow,g_\downarrow)
 +\delta_\theta\norm\eta^2.
 \label{eq:symmetrized-square-bound}
\end{align}
\end{proposition}

\begin{proof}
Since $\mathbb H\varphi=E^{\F}(2,Z)\varphi$ and $\mathbb W a_\uparrow(f)\varphi=0$, we have
\begin{align*}
 &\ip{a_\uparrow^*(f)\varphi}
              {(\mathbb H-E^{\F}(2,Z))a_\uparrow^*(f)\varphi}\\
 &=\ip\varphi{a_\uparrow(f)
                         [\mathbb H,a_\uparrow^*(f)]\varphi}\\
 &=\ip\varphi{
       \{a_\uparrow(f),[\mathbb H,a_\uparrow^*(f)]\}\varphi}
       -\ip\varphi{[\mathbb H,a_\uparrow^*(f)]a_\uparrow(f)\varphi}\\
 &=\ip f{(h_Z+2D_{\gamma_\varphi}-K_{\gamma_\varphi})f}
       -\ip{\mathbb H\varphi}{a_\uparrow^*(f)a_\uparrow(f)\varphi}
       +\ip{a_\uparrow(f)\varphi}{\mathbb H a_\uparrow(f)\varphi}\\
 &=\ip f{(h_Z+2D_{\gamma_\varphi}-K_{\gamma_\varphi})f}
       +\ip{a_\uparrow(f)\varphi}{(\mathbb H-E^{\F}(2,Z))a_\uparrow(f)\varphi}\\
 &=\ip f{(h_Z+2D_{\gamma_\varphi}-K_{\gamma_\varphi})f}
       +\ip{\mathcal R_\varphi\overline f}
                         {(h_Z-E^{\F}(2,Z))\mathcal R_\varphi\overline f}\\
 &=\ip f{(h_Z+2D_{\gamma_\varphi}-K_{\gamma_\varphi})f}
       +\overline{\ip{\mathcal R_\varphi f}
                         {(h_Z-E^{\F}(2,Z))\mathcal R_\varphi f}}\\
 &=\ip f{(h_Z+2D_{\gamma_\varphi}-K_{\gamma_\varphi})f}
       +\ip{\mathcal R_\varphi f}{(h_Z-E^{\F}(2,Z))\mathcal R_\varphi f}.
\end{align*}
The last two equalities follow because $h_Z$ and $\mathcal R_\varphi$ are real and the quadratic form of $h_Z-E^{\F}(2,Z)$ is real-valued. This proves \eqref{eq:exact-hole}. Combining \eqref{eq:exact-hole} with Proposition~\ref{prop:completion-of-the-square} gives \eqref{eq:symmetrized-square-bound}.
\end{proof}

Define
\begin{align}
 S_{\Gamma_\theta}(f)
 :=\inf_{\lambda\in\C}\Bigl\{
 \bigl\|A_{\Gamma_\theta}^{-1/2}
       \bigl((\tr\Gamma_\theta)f-\lambda e_0\bigr)\bigr\|^2
       +\bigl\|(A_{\Gamma_\theta}-L_{\Gamma_\theta})^{-1/2}
       \bigl(((\tr\Gamma_\theta)\1-\Gamma_\theta)f
                   -\lambda e_0\bigr)\bigr\|^2\Bigr\}.
 \label{eq:source-inverse-definition}
\end{align}
The inverse quadratic forms take the value $+\infty$ outside their domains.

\begin{lemma}\label{lem:source-variational}
 For $S_{\Gamma_\theta}(f)$ as in \eqref{eq:source-inverse-definition}, we have
 \begin{equation}\label{eq:source-variational}
      \inf_{g_\uparrow+g_\downarrow\perp e_0}\mathcal C_{\Gamma_\theta}(f;g_\uparrow,g_\downarrow)=\ip{f}{(2D_{\Gamma_\theta}-K_{\Gamma_\theta})f}-S_{\Gamma_\theta}(f),
 \end{equation}
 If $\psi\in H^1\cap\mathcal K_{2,1}$ has the decomposition \eqref{eq:remainder-conditions}, then
 \begin{equation}\label{eq:source-variational-lower-bound-quadratic-F}
      \ip\psi{(H_{3,Z}^{\mathrm{sc}}-E^{\F}(2,Z))\psi}\geq\mathfrak F_\theta[f]+\delta_\theta\norm\eta^2,
 \end{equation}
 where
 \begin{equation}\label{eq:def-of-quadratic-F}
      \mathfrak F_\theta[f]:=\ip f{h_Zf}+\ip{\mathcal R_\varphi f}{(h_Z-E^{\F}(2,Z))\mathcal R_\varphi f}-\frac{\theta}{2-\theta}\ip{f}{(2D_{\Gamma_{+,\phi}}-K_{\Gamma_{+,\phi}})f}+S_{\Gamma_\theta}(f).
 \end{equation}
\end{lemma}

\begin{proof}[Proof of Lemma~\ref{lem:source-variational}]
For bounded, uniformly positive operators, the Euler equations for the constrained minimum in \eqref{eq:source-variational} are
\begin{equation}\label{eq:source-euler-equations}
 \begin{split}
 A_{\Gamma_\theta}g_\uparrow
   &=(\tr\Gamma_\theta)f-\lambda e_0,\\
 (A_{\Gamma_\theta}-L_{\Gamma_\theta})g_\downarrow
   &=((\tr\Gamma_\theta)\1-\Gamma_\theta)f-\lambda e_0,\\
 \ip{e_0}{g_\uparrow+g_\downarrow}&=0.
 \end{split}
\end{equation}
Completing both squares gives
\begin{align*}
 \inf_{g_\uparrow+g_\downarrow\perp e_0}
       \mathcal C_{\Gamma_\theta}(f;g_\uparrow,g_\downarrow)
 &=\ip f{(2D_{\Gamma_\theta}-K_{\Gamma_\theta})f}
       -\inf_{\lambda\in\C}\Bigl\{
       \bigl\|A_{\Gamma_\theta}^{-1/2}
       \bigl((\tr\Gamma_\theta)f-\lambda e_0\bigr)\bigr\|^2\\
 &\qquad
       +\bigl\|(A_{\Gamma_\theta}-L_{\Gamma_\theta})^{-1/2}
       \bigl(((\tr\Gamma_\theta)\1-\Gamma_\theta)f
                    -\lambda e_0\bigr)\bigr\|^2\Bigr\}\\
 &=\ip f{(2D_{\Gamma_\theta}-K_{\Gamma_\theta})f}
       -S_{\Gamma_\theta}(f).
\end{align*}
Substituting this identity and \eqref{eq:source-density} into \eqref{eq:symmetrized-square-bound} proves \eqref{eq:source-variational-lower-bound-quadratic-F}.
\end{proof}

\section{Scalar reduction}\label{sec:scalar}

From now on, $Z=2$. Let $\varphi$ be the normalized positive ground state of $H_{2,2}^{\mathrm{sc}}$ on $\mathcal K_{1,1}$ with energy $E^{\F}(2,2)$. The densities $\gamma_\varphi$ and $\Gamma_{+,\varphi}$ are defined in \eqref{eq:core-density} and \eqref{eq:source-density}, respectively.

\begin{proposition}\label{prop:scalar-source-reduction}
For every $0<\theta\leq2/3$ and every $f\in H^1(\mathbb R^3)\cap e_0^\perp$, we have
\begin{equation}\label{eq:scalar-source-reduction}
\begin{aligned}
    \mathfrak F_\theta[f]
    \geq
    \frac12\|\nabla f\|_2^2
    -\tr{\Gamma_\theta}
    \int_{\mathbb R^3}
    \Bigl(
        \frac2{|x|}
        -\frac2{\sqrt{|x|^2+2}}
    \Bigr)|f(x)|^2\,dx
    -\frac{\tr\Gamma_\theta}{10}
 \int_{\R^3}
 \frac{|P_{\mathrm{odd}}f|^2}{(|x|^2+2)^{3/2}}\dd x,
\end{aligned}
\end{equation}
where $\mathfrak F_\theta$ is defined in \eqref{eq:def-of-quadratic-F} and $P_{\mathrm{odd}}f=\frac{1}{2}(f(x)-f(-x))$.
\end{proposition}

\begin{proof}
By \eqref{eq:source-inverse-definition}, \eqref{eq:source-density}, and the operator Cauchy--Schwarz inequality, we have
\begin{align}
 &\langle\mathcal R_\varphi f,(h_2-E^{\F}(2,2))\mathcal R_\varphi f\rangle
       +S_{\Gamma_\theta}(f)
       \geq\inf_{\lambda\in\C}\big\langle
       (\tr\Gamma_\theta)f-\lambda e_0,\,
       (X^{-1}+Y^{-1})
       \bigl((\tr\Gamma_\theta)f-\lambda e_0\bigr)
       \big\rangle,
 \label{eq:scalar-proof-direct-inverses}
\end{align}
where
$$
\begin{aligned}
X=A_{\Gamma_\theta}\quad,\qquad
Y=A_{\Gamma_\theta}-L_{\Gamma_\theta}
+\mathcal R_\varphi
\Big(P_0+\frac2{2-\theta}Q_0\Big)
(h_2-E^{\F}(2,2))^{-1}
\Big(P_0+\frac2{2-\theta}Q_0\Big)
\mathcal R_\varphi.
\end{aligned}
$$
Spectral calculus and convexity of inversion give
\begin{align}
 &\inf_{\lambda\in\C}
  \langle f-\lambda e_0,(X^{-1}+Y^{-1})(f-\lambda e_0)\rangle
  \notag\\
 &\quad=\frac1\pi\inf_{\lambda\in\C}
  \int_0^\infty t^{-1/2}
  \Bigl\langle f-\lambda e_0,
   \bigl((X^2+t)^{-1}+(Y^2+t)^{-1}\bigr)
   (f-\lambda e_0)\Bigr\rangle\dd t\notag\\
 &\quad\geq\frac4\pi\int_0^\infty t^{-1/2}
  \inf_{\lambda\in\C}
  \langle f-\lambda e_0,
       (X^2+Y^2+2t)^{-1}(f-\lambda e_0)\rangle\dd t
  \notag\\
 &\quad=\frac4\pi\int_0^\infty t^{-1/2}
  \Bigl\langle f,
    \bigl(Q_0(X^2+Y^2)Q_0+2tQ_0\bigr)^{-1}f
  \Bigr\rangle\dd t\notag\\
 &\quad=2\Big\langle f,
    \Bigl(\frac12Q_0(X^2+Y^2)Q_0\Bigr)^{-1/2}f
  \Big\rangle.
 \label{eq:scalar-proof-resolvent-combination}
\end{align}
Combining \eqref{eq:scalar-proof-direct-inverses} and \eqref{eq:scalar-proof-resolvent-combination}, we obtain
\begin{align}
 \langle\mathcal R_\varphi f,(h_2-E^{\F}(2,2))\mathcal R_\varphi f\rangle
 +S_{\Gamma_\theta}(f)
 \geq
 2(\tr\Gamma_\theta)^2\Big\langle f,
 \Big(\frac12Q_0(X^2+Y^2)Q_0
 \Big)^{-1/2}f\Big\rangle.
 \label{eq:scalar-proof-inverse}
\end{align}
The inverses act on $Q_0\mathfrak h$.

\noindent{\textbf{Claim.}} The following quadratic form inequality holds on $Q_0\mathfrak h$.
\begin{equation}\label{eq:claim-quadratic-ineq}
      X^2+Y^2\leq2(\tr{\Gamma_\theta})^2\Big(|\cdot|^2+\frac{5}{3}+\frac{\tr(|\cdot|^2\Gamma_{0,\phi})}{3\tr{\Gamma_\theta}}P_{\mathrm{odd}}\Big).
\end{equation}
For normalized $v$ in this space, we have
\begin{align}
&\frac{\ip{v}{X^2v}+\ip{v}{Y^2v}}{\operatorname{Tr}\Gamma_\theta}
 -2\operatorname{Tr}\Gamma_\theta\,\ip{v}{|\cdot|^2v}
 -\frac23\operatorname{Tr}(|\cdot|^2\Gamma_{0,\phi})
       \|P_{\mathrm{odd}}v\|^2
\nn\\
&\le
\operatorname{Tr}(|\cdot|^2\gamma_\phi)
+\Big(\frac6{2-\theta}-1\Big)
       \operatorname{Tr}(|\cdot|^2\Gamma_{+,\phi})
\nn\\
&+
\operatorname{Tr}(|\cdot|^2\Gamma_\theta)
-2\operatorname{Re}\langle |\cdot|^2v,\Gamma_\theta v\rangle
+\Big\|
 \Big(P_0+\frac2{2-\theta}Q_0\Big)^{3/2}
 (h_2-E^{\F}(2,2))^{-1}\mathcal R_\phi v
\Big\|^2
\nn\\
&\le
\operatorname{Tr}(|\cdot|^2\gamma_\phi)
+\Big(\frac6{2-\theta}-1\Big)
       \operatorname{Tr}(|\cdot|^2\Gamma_{+,\phi})
+\operatorname{Tr}\Gamma_\theta
       \langle e_0,|\cdot|^2e_0\rangle
\nn\\
&+
\sqrt{\frac2{2-\theta}\operatorname{Tr}\Gamma_{+,\phi}}\,
\Big(
 \|Q_0|\cdot|^2e_0\|
 +\sqrt{\frac2{2-\theta}}\,
 \left\|
 \bigl(|\cdot|^2-\langle e_0,|\cdot|^2e_0\rangle\bigr)
 \mathcal R_\phi
 \right\|_{\mathrm{HS}}
\Big)
\nn\\
&\le
\operatorname{Tr}(|\cdot|^2\gamma_\phi)
+\Big(\frac6{2-\theta}-1\Big)
       \operatorname{Tr}(|\cdot|^2\Gamma_{+,\phi})
+\operatorname{Tr}\Gamma_\theta
       \langle e_0,|\cdot|^2e_0\rangle
\nn\\
&+
\sqrt{\frac2{2-\theta}\operatorname{Tr}\Gamma_{+,\phi}}\,
\Big(
 \|Q_0|\cdot|^2e_0\|
 +\sqrt{\frac2{2-\theta}\Big(\frac{2\operatorname{Tr}(|\cdot|^2\gamma_\phi)}{\mu-E^{\F}(2,2)}
+\left(
 \operatorname{Tr}(|\cdot|^2\gamma_\phi)
 -\langle e_0,|\cdot|^2e_0\rangle
\right)^2\Big)}
\Big).\label{eq:Neu}
\end{align}
Proposition~\ref{prop:appendix-density-moments} and $\tr(\Gamma_\theta)=1+\frac{\theta}{2-\theta}\tr(\Gamma_{+,\phi})$ give
\begin{align*}
      \mathrm{RHS\ of}\ \eqref{eq:Neu}-\frac{10}{3}\tr(\Gamma_\theta)\leq-0.996+4.125\sqrt{\tr{\Gamma_{+,\phi}}}-1.291\tr(\Gamma_{+,\phi})\leq-0.0454375<0,
\end{align*}
which proves \eqref{eq:claim-quadratic-ineq}. Combining this estimate with \eqref{eq:scalar-proof-inverse} gives
\begin{align}
 \langle\mathcal R_\varphi f,(h_2-E^{\F}(2,2))\mathcal R_\varphi f\rangle
 +S_{\Gamma_\theta}(f)
 &\geq
 2(\tr\Gamma_\theta)\Big\langle f,
 \Big(Q_0|\cdot|^2Q_0+\frac{5}{3}Q_0+\frac{\tr(|\cdot|^2\Gamma_{0,\phi})}{3\tr{\Gamma_\theta}}P_{\mathrm{odd}}
 \Big)^{-1/2}f\Big\rangle\nn\\
 &\geq
 2(\tr\Gamma_\theta)\Big\langle f,
 \Big(|\cdot|^2+2+\frac{1}{10}P_{\mathrm{odd}}
 \Big)^{-1/2}f\Big\rangle,\nn
\end{align}
where the first inequality uses $Q_0P_{\mathrm{odd}}Q_0\leq P_{\mathrm{odd}}$, which follows from $[P_{\mathrm{odd}},Q_0]=0$. For the second inequality, we complete the square and use the bounds from Proposition~\ref{prop:appendix-density-moments},
$$
 \frac{\tr(|\cdot|^2\Gamma_{0,\varphi})}
      {3\tr\Gamma_\theta}\leq\frac{\tr(|\cdot|^2\gamma_\phi)}{3}\leq\frac{13}{30},
 \qquad
 2+\langle e_0,|\cdot|^2e_0\rangle
      -3\|Q_0|\cdot|^2e_0\|^2=\frac{7}{32}>0.
$$
Using $P_{\mathrm{odd}}\prec Q_0$ and the orthogonal decomposition $\1=P_0+(Q_0-P_{\mathrm{odd}})+P_{\mathrm{odd}}$, we obtain
$$
\begin{aligned}
&|\cdot|^2+2+\frac1{10}P_{\mathrm{odd}}-\big(Q_0|\cdot|^2Q_0+\frac53Q_0+\frac{\tr(|\cdot|^2\Gamma_{0,\phi})}{3\tr{\Gamma_\theta}}P_{\mathrm{odd}}\big)\\
&=
\Big(\frac{13}{30}-\frac{\tr(|\cdot|^2\Gamma_{0,\phi})}{3\tr{\Gamma_\theta}}\Big)P_{\mathrm{odd}}+
\frac13
\bigl(Q_0-P_{\mathrm{odd}}+3Q_0|\cdot|^2P_0\bigr)^*
\bigl(Q_0-P_{\mathrm{odd}}+3Q_0|\cdot|^2P_0\bigr)\\
&+
\left(
2+\langle e_0,|x|^2e_0\rangle
-3\|Q_0|x|^2e_0\|^2
\right)P_0\geq0.
\end{aligned}
$$
Since $(a+d)^{-1/2}\geq a^{-1/2}-d/(2a^{3/2})$, we obtain
\begin{align*}
 &\langle\mathcal R_\varphi f,
          (h_2-E^{\F}(2,2))\mathcal R_\varphi f\rangle
       +S_{\Gamma_\theta}(f)\geq
 2(\tr\Gamma_\theta)
 \int_{\R^3}\frac{|f(x)|^2}{\sqrt{|x|^2+2}}\dd x
 -\frac{\tr\Gamma_\theta}{10}
 \int_{\R^3}
 \frac{|P_{\mathrm{odd}}f|^2}{(|x|^2+2)^{3/2}}\dd x.
\end{align*}
Finally, positivity of the Coulomb form and Newton's formula give
$$
 K_{\Gamma_{+,\varphi}}\geq0,
 \qquad
 D_{\Gamma_{+,\varphi}}(x)
 \leq\frac{\tr\Gamma_{+,\varphi}}{|x|}.
$$
Inserting these bounds into \eqref{eq:def-of-quadratic-F} gives
\begin{align*}
 \mathfrak F_\theta[f]
 &=\frac12\|\nabla f\|_2^2
   -2\int_{\R^3}\frac{|f(x)|^2}{|x|}\dd x
   +\langle\mathcal R_\varphi f,
        (h_2-E^{\F}(2,2))\mathcal R_\varphi f\rangle+S_{\Gamma_\theta}(f)\\
 &\qquad
   -\frac{2\theta}{2-\theta}
       \langle f,D_{\Gamma_{+,\varphi}}f\rangle
   +\frac{\theta}{2-\theta}
       \langle f,K_{\Gamma_{+,\varphi}}f\rangle\\
 &\geq\frac12\|\nabla f\|_2^2
 -2(\tr{\Gamma_\theta})
       \int_{\R^3}\frac{|f(x)|^2}{|x|}\dd x+2(\tr\Gamma_\theta)
       \int_{\R^3}\frac{|f(x)|^2}{\sqrt{|x|^2+2}}\dd x
       -\frac{\tr\Gamma_\theta}{10}
 \int_{\R^3}
 \frac{|P_{\mathrm{odd}}f|^2}{(|x|^2+2)^{3/2}}\dd x\\
 &=\frac12\|\nabla f\|_2^2
 -\tr{\Gamma_\theta}
       \int_{\R^3}\Bigl(\frac2{|x|}
             -\frac2{\sqrt{|x|^2+2}}\Bigr)|f(x)|^2\dd x
       -\frac{\tr\Gamma_\theta}{10}
 \int_{\R^3}
 \frac{|P_{\mathrm{odd}}f|^2}{(|x|^2+2)^{3/2}}\dd x.
\end{align*}
\end{proof}

\section{One-dimensional variational problem and proof of coercivity}\label{sec:1-d-variational}

To absorb the nonpositive terms into the kinetic term, we use the bound from \eqref{eq:scalar-source-reduction}
\begin{equation}
      \mathfrak F_\theta[f]\geq\frac{1-4C_*\tr(\Gamma_\theta)}{2}\norm{\nabla f}_2^2,\quad\forall f\in H^1(\mathbb R^3)\cap e_0^\perp,\ \theta\in(0,\frac{2}{3}],
\end{equation}
where
\begin{equation}\label{eq:def-of-C-star}
C_*=\sup_{\substack{
f\perp e_0,\norm{\nabla f}_2^2=1
}}\Big\{\,\int_{\R^3}V|f|^2\dd x+\frac{1}{20}\int_{\R^3}\frac{|P_{\mathrm{odd}}f|^2}{(|x|^2+2)^{3/2}}\dd x\,\Big\},\qquad V(r):=\frac{1}{r}-\frac{1}{\sqrt{r^2+2}}.
\end{equation}
We bound $C_*$ from above to obtain $\mathfrak F_\theta[f]\gtrsim\norm{\nabla f}_2^2$, then choose $\theta$ and apply Lemma~\ref{lem:source-variational} to prove Theorem~\ref{thm:main-reduced}.

The supremum in \eqref{eq:def-of-C-star} reduces to a one-dimensional variational problem in which the odd term vanishes.
\begin{lemma}\label{lem:reduced-C-star}
We have
$$
C_*=
\sup_{\substack{
0\ne u\in \dot H^1_0(0,\infty),\\
\int_0^\infty r e^{-2r}u(r)\,dr=0
}}
\frac{
\displaystyle\int_0^\infty
V(r)|u(r)|^2\,dr
}{
\displaystyle\int_0^\infty|u'(r)|^2\,dr
}.
$$
\end{lemma}

\begin{proof}
Let $C_{\mathrm{rad}}$ and $C_{\mathrm{nr}}$ denote the suprema in \eqref{eq:def-of-C-star} over radial functions and over functions with zero spherical average, respectively. It suffices to show that $C_*=C_{\mathrm{rad}}$. Since $e_0$ is radial and $P_{\rm odd}f=P_{\rm odd}f_{\rm nr}$, we have $C_*=\max\{C_{\rm rad},C_{\rm nr}\}$.

On the one hand, an elementary calculation gives
$$
 V(r)+\frac1{20(r^2+2)^{3/2}}\leq\frac{15}{8r(r+1)^2}.
$$
For $f$ with zero spherical average $\int_{\mathbb S^2}f(r\omega)\dd\omega=0$ for almost every $r>0$, we have
\begin{align*}
 &\qquad\int_{\R^3}V|f|^2\dd x
 +\frac1{20}\int_{\R^3}
       \frac{|P_{\mathrm{odd}}f|^2}{(|x|^2+2)^{3/2}}\dd x\\
 &\leq\int_{\R^3}V|f|^2\dd x
 +\frac1{20}\int_{\R^3}
       \frac{|f|^2}{(|x|^2+2)^{3/2}}\dd x
 \leq\frac{15}{8}\int_{\R^3}\frac{|f(x)|^2}{|x|(|x|+1)^2}\dd x\leq\frac{5}{32}\|\nabla f\|_2^2,
\end{align*}
where the last inequality follows from the Hardy-type inequality and the first nonzero eigenvalue $2$ of $-\Delta_{\mathbb S^2}$. Indeed,
$$
\begin{aligned}
\norm{\nabla f}_2^2
-12\int_{\mathbb R^3}\frac{|f|^2}{|x|(|x|+1)^2}\dd x
=
\int_0^\infty\!\!\int_{\mathbb S^2}
r^2h^2
\left|\partial_r\!\left(\frac fh\right)\right|^2
\,d\omega\,dr+\int_0^\infty\!\!\int_{\mathbb S^2}
\bigl(|\nabla_\omega f|^2-2|f|^2\bigr)
\,d\omega\,dr\ge0,
\end{aligned}
$$
where $h(r)=r/(r+1)^3$. Thus, $C_{\rm nr}\leq 5/32$.

On the other hand, by a density argument, we may test \eqref{eq:def-of-C-star} with the radial function
$F(r)=\frac{1-(1+6r)e^{-2r}}{r}\in\dot H^1(\mathbb R^3)$, which satisfies $\int_{\mathbb R^3}F(x)e_0(x)\,dx=0$.
This gives
$$
\begin{aligned}
C_{\rm rad}
&\ge
\frac25\int_0^\infty
V(r)\bigl[1-(1+6r)e^{-2r}\bigr]^2\,dr\\
&\ge
\frac25\int_0^\infty
\frac{e^{-r}}r
\bigl[1-(1+6r)e^{-2r}\bigr]^2\,dr\\
&=
\frac25\left(\log\frac95-\frac4{25}\right)
>\frac{5}{32},
\end{aligned}
$$
so the radial supremum dominates the nonradial part. This concludes the proof.
\end{proof}

Before estimating $C_*$, we introduce some notations. For a positive weight function $W(r)$ on $(0,\infty)$ and $\alpha>0$, define
\begin{equation}\label{eq:def-of-M-A}
      M_W(\alpha):=\sup_{\substack{
0\ne u\in \dot H_0^1(0,\infty),\\
\int_0^\infty r e^{-\alpha r}u(r)\,dr=0
}}
\frac{
\displaystyle\int_0^\infty
W(r)|u(r)|^2\,dr
}{
\displaystyle\int_0^\infty|u'(r)|^2\,dr
}\in [0,\infty].
\end{equation}
Then $C_*=M_V(2)$, and a change of variables gives
\begin{equation}\label{eq:rescale}
      M_{W_b}(\alpha)=M_W(b\alpha),\qquad W_b(r):=\frac{1}{b^2}W\big(\frac{r}{b}\big),\qquad b>0.
\end{equation}
The idea is conceptual simple. To give an upper-bound of $C_*$, we will choose a weight $W$ for which \eqref{eq:def-of-M-A} can be solved explicitly. By \eqref{eq:rescale}, this gives
$$
 C_*=M_V(2)\leq \inf_{b>0}\Big(M_W(2b)\cdot\sup_{r>0}\frac{V(r)}{W_b(r)}\Big).
$$
For suitable $W$, the variational problem \eqref{eq:def-of-M-A} is related to the second eigenvalue of the self-adjoint operator formally given by $-W^{-1}\frac{\d^2}{\dd r^2}$. We use the min--max formula
$$
    \lambda_2(W)
    :=
    \inf_{\substack{E\subset\dot H_0^1(0,\infty)\\ \dim E=2}}
    \ \sup_{u\in E\setminus\{0\}}
    \frac{\displaystyle\int_0^\infty |u'(r)|^2\,dr}
         {\displaystyle\int_0^\infty W(r)|u(r)|^2\,dr}.
$$

\begin{lemma}
\label{lem:log-convex-weight}
Suppose that $W\in C(0,\infty)$ is positive and $\int_0^\infty rW(r)\dd r<\infty$. Let $v\in\dot H_0^1(0,\infty)\setminus\{0\}$ be a real-valued eigenfunction associated with $\lambda_2(W)$, so that
$$
    -v''=\lambda_2(W)Wv
    \qquad\text{on }(0,\infty).
$$
Then $v$ has a unique simple zero $a$ in $(0,\infty)$. Choose the sign of $v$ so that $v'(a)>0$ and assume that
$$
    r\longmapsto
    \frac{v(r)-(r-a)v'(r)}{r(r-a)^3}
$$
is log-convex, with the value at $r=a$ defined by continuity. Then there exists a unique $\alpha_W>0$ such that
\begin{equation}\label{eq:def-of-alpha-A}
    \int_0^\infty r e^{-\alpha_W r}v(r)\,dr=0.
\end{equation}
Moreover, $M_W(\alpha_W)=\frac1{\lambda_2(W)}.$
\end{lemma}

\begin{proof}
By Sturm's oscillation theorem \cite[Theorem~2.1]{GST96}, $v$ has a unique simple zero $a$ in $(0,\infty)$. Then $v<0$ on $(0,a)$ and $v>0$ on $(a,\infty)$. The quadratic form
$$
 \int_0^\infty|u'|^2\dd r
 -\lambda_2(W)\int_0^\infty W|u|^2\dd r
$$
has negative index one. On $(a,\infty)$, $v$ is positive, concave, and increasing. Hence the moment in \eqref{eq:def-of-alpha-A} is positive for small $\alpha$. A fixed interval to the left of $a$ on which $v<0$ and the bound $|v(r)|\le\sqrt r\,\|v'\|_2$ give a negative moment for large $\alpha$. Since
$$
 \frac{\dd}{\dd\alpha}\Bigl(
 e^{a\alpha}\int_0^\infty re^{-\alpha r}v(r)\dd r\Bigr)
 =\int_0^\infty r(a-r)e^{-\alpha(r-a)}v(r)\dd r<0,
$$
the zero $\alpha_W$ is unique.

We set
$$
 z(r)=\frac{v(r)}{r-a},\qquad
 G(r)=-\frac{2z'(r)}{r-a},\qquad
 g(r)=re^{-\alpha_Wr}.
$$
The equation and the signs of $v$ give $z,G>0$, with continuous values at $a$, and
$$
 z''+\lambda_2(W)Wz=G,\qquad Gv=-(z^2)'.
$$
As $z(0)=z(\infty)=0$, we have $\int_0^\infty Gv=0$. Choose $c>0$ so that
$$
 \int_0^a(g-cG)v\dd r=0,\qquad
 H(r)=\int_0^r(g-cG)v\dd s.
$$
Then $H(0)=H(a)=H(\infty)=0$. By assumption, $g/G$ is log-concave. Its weighted averages with respect to $G|v|\dd r$ on the two nodal intervals both equal $c$. Thus $g-cG$ has signs $-,+,-$ in that order on possibly degenerate intervals, with $a$ in the middle interval. The signs of $v$ then give $H\ge0$.

Define
$$
 Z(r)=cz(r)+v(r)\int_a^r\frac{H(s)}{v(s)^2}\dd s\ge cz(r)>0.
$$
The integrand extends continuously through $a$, since $H(a)=H'(a)=0$. Moreover,
$$
 \frac{H(r)}{v(r)^2}
 =\frac{c}{(r-a)^2}
  +\frac{\int_0^r gv\dd s}{v(r)^2}
$$
is bounded near zero and is $O(r^{-2})$ at infinity. Near zero, concavity of $-v$ gives $|v(r)|\gtrsim r$, and $|v(r)|\le\sqrt r\,\|v'\|_2$. At infinity, $v$ is bounded below and $\int_0^r gv=-\int_r^\infty gv$ decays exponentially. Thus $Z\in\dot H_0^1(0,\infty)$, and
\begin{align*}
 -Z''-\lambda_2(W)WZ
 &=-cG-\frac1v\Bigl(v^2\frac{H}{v^2}\Bigr)'=-cG-\frac{H'}v=-g.
\end{align*}
It follows that
$$
 \int_0^\infty\bigl(|Z'|^2-\lambda_2(W)WZ^2\bigr)\dd r
 =-\int_0^\infty gZ\dd r<0.
$$
For every real $u$ satisfying $\int_0^\infty gu=0$, we also have
$$
 \int_0^\infty\bigl(u'Z'-\lambda_2(W)WuZ\bigr)\dd r=0.
$$
Since the quadratic form has negative index one, it is nonnegative on this constraint space. Applying this to the real and imaginary parts gives nonnegativity for complex $u$. Testing with $v$ proves $M_W(\alpha_W)=1/\lambda_2(W)$.
\end{proof}

We choose
\begin{equation}\label{eq:def-of-A-and-v}
W(r)=
\frac{24(3r^2+7r+3)}
{r(4r^2+5r+4)^2},
\qquad
v(r)=
\frac{4r(r-1)}{4r^2+5r+4}.
\end{equation}
An elementary calculation gives
\begin{equation}\label{eq:property-A-and-v}
      \Big(\log \frac{v(r)-(r-1)v'(r)}{r(r-1)^3}\Big)''>0;\quad\int_0^\infty rW(r)\dd r<\infty;\quad -v''=Wv,
\end{equation}
and
\begin{equation}\label{eq:comparisom-A-V}
      W_b(r)\geq\min\Big\{\frac{9}{2b},\frac{25}{6}\Big\}V(r)\quad,\quad\forall\  b\geq1\ , \ r>0.
\end{equation}

\begin{corollary}\label{cor:uniform-variational-coercivity}
We have $C_*\leq \frac{6}{25}$. In particular, for every $0<\theta\leq2/3$ and every $f\in H^1(\mathbb R^3)\cap e_0^\perp$,
\begin{equation}\label{eq:kinetic-variational-coercivity}
 \mathfrak F_\theta[f]
 \geq\frac{1}{200}\|\nabla f\|_2^2.
\end{equation}
\end{corollary}

\begin{proof}
Let $W$ and $v$ be given by \eqref{eq:def-of-A-and-v}. Since $v$ has one zero in $(0,\infty)$, Sturm's oscillation theorem \cite[Theorem~2.1]{GST96} and $-v''=Wv$ give $\lambda_2(W)=1$. Evaluating the moment in \eqref{eq:def-of-alpha-A} of Lemma~\ref{lem:log-convex-weight} gives $2<\alpha_W<15/7$.

Taking $b_0=\alpha_W/2$ and applying Lemmas~\ref{lem:log-convex-weight} and~\ref{lem:reduced-C-star} with \eqref{eq:comparisom-A-V}, we obtain
$$
 C_*=M_V(2)\leq M_W(2b_0)\cdot\sup_{r>0}\frac{V(r)}{W_{b_0}(r)}\leq\max\Big\{\frac{2b_0}{9},\frac{6}{25}\Big\}=\frac{6}{25},
$$
where the last equality uses $2b_0=\alpha_W<15/7$.

By \eqref{eq:def-of-C-star} and Proposition~\ref{prop:scalar-source-reduction},
$$
 \mathfrak F_\theta[f]\geq\frac{1-4C_*\tr(\Gamma_\theta)}{2}\norm{\nabla f}_2^2\geq\frac{1}{200}\norm{\nabla f}_2^2,\quad\forall f\in H^1(\mathbb R^3)\cap e_0^\perp,\ \theta\in(0,\frac{2}{3}],
$$
where $\tr(\Gamma_\theta)=1+\frac{\theta}{2-\theta}\tr(\Gamma_{+,\phi})\leq1+\frac{1}{2}\tr(\Gamma_{+,\phi})\leq\frac{33}{32}$ by Proposition~\ref{prop:appendix-density-moments}.
\end{proof}

\begin{proof}[Proof of Theorem~\ref{thm:main-reduced}]
We take $\mu=-35/16$ and $\varepsilon_0=19/200$ by Proposition~\ref{prop:appendix-energy-overlap} and set $\theta=2/3$. Since $\lambda_1=-2$ and $\lambda_2=-1/2$, these parameters satisfy the hypotheses of Proposition~\ref{prop:remainder}. Equations \eqref{eq:remainder-coefficient} and \eqref{eq:appendix-energy-bounds} give
$$
 \delta_\theta=-E^{\F}(2,2)-\frac{9269}{3200}>\frac1{200}.
$$
For $\psi\in H^1\cap\mathcal K_{2,1}$, use the decomposition $\psi=a_\uparrow^*(f)\varphi+\eta$ from \eqref{eq:remainder-conditions}. Then \eqref{eq:source-variational-lower-bound-quadratic-F} and Corollary~\ref{cor:uniform-variational-coercivity} give
$$
 \ip\psi{(H_{3,2}^{\mathrm{sc}}-E^{\F}(2,2))\psi}
 \geq\mathfrak F_\theta[f]+\delta_\theta\norm\eta^2
 \geq\frac1{200}\norm{\nabla f}_2^2+\frac1{200}\norm\eta^2.
$$
This proves Theorem~\ref{thm:main-reduced}.
\end{proof}

\appendix

\section{Estimates for the helium atom}\label{app:core-inputs}

Let $\varphi$ be the normalized positive ground state of $H_{2,2}^{\mathrm{sc}}$ on $\mathcal K_{1,1}$, with energy $E^{\F}(2,2)$. We use the operators and densities defined in \eqref{eq:core-density} and \eqref{eq:source-density}.

\subsection{Excited spectrum}\label{app:excited-spectrum}

\begin{proposition}\label{prop:appendix-spectral-gap}
We have the quadratic-form bound
\begin{equation}\label{eq:appendix-spectral-gap}
 H_{2,2}^{\mathrm{sc}}\big|_{\varphi^\perp}
 \geq-\frac{35}{16}
 \qquad\text{on }\mathcal K_{1,1}=L^2(\R^6).
\end{equation}
\end{proposition}

\begin{proof}
For $n=2,3$, let
$$
 \mathcal P_n
 =P_0\otimes\1_{\{\lambda_n\}}(h_2)
  +\1_{\{\lambda_n\}}(h_2)\otimes P_0.
$$
Multiplying the radial equation in \eqref{eq:one-particle-angular-sum} by $r^{k+1}R'_{n,\ell}+(k+1)r^kR_{n,\ell}/2$ and integrating by parts gives
\begin{align*}
 &2(k+1)\lambda_n\int_0^\infty r^k|R_{n,\ell}|^2\dd r
 +2(2k+1)\int_0^\infty r^{k-1}|R_{n,\ell}|^2\dd r\\
 &\qquad
 +\frac{k}{4}\bigl(k^2-(2\ell+1)^2\bigr)
       \int_0^\infty r^{k-2}|R_{n,\ell}|^2\dd r=0
 \qquad(k=0,1,2).
\end{align*}
Substituting $\lambda_n=-2/n^2$ and applying the recurrence yields
$$
 \int_0^\infty r^2|R_{n,\ell}(r)|^2\dd r
 =\frac{n^2}{8}\bigl(5n^2+1-3\ell(\ell+1)\bigr).
$$
In the compression of $|x-y|^2$, the term $-2x\cdot y$ couples the two summands of $\mathcal P_n$ only at $\ell=1$. The coupling has norm at most $2\|x_je_0\|_2^2=1/2$. The radial moment at $\ell=1$ is smaller than at $\ell=0$ by $3n^2/4$. Thus
\begin{equation}\label{eq:appendix-shell-second-moments}
 \mathcal P_2|x-y|^2\mathcal P_2\leq\frac{45}{4}\mathcal P_2,
 \qquad
 \mathcal P_3|x-y|^2\mathcal P_3\leq\frac{105}{2}\mathcal P_3.
\end{equation}
The $\ell=1$ part of the first compression is bounded by $35/4$.

For $\psi_{2,0,0}(x)=\pi^{-1/2}(1-|x|)e^{-|x|}$, spherical integration gives
$$
 A_{P_0}(r)
 =r+\frac1{4r}-\Bigl(\frac1{4r}+\frac14\Bigr)e^{-4r},
 \qquad
 \bigl(|\cdot|^{-1}*(e_0\psi_{2,0,0})\bigr)(r)
 =\frac{8\sqrt2}{27}(3r+1)e^{-3r}.
$$
Consequently,
$$
 \langle e_0\otimes\psi_{2,0,0},
       |x-y|\,e_0\otimes\psi_{2,0,0}\rangle
 =\langle\psi_{2,0,0},A_{P_0}\psi_{2,0,0}\rangle
 =\frac{1007}{324}.
$$
Likewise,
$$
 \langle e_0\otimes\psi_{2,0,0},
       |x-y|\,\psi_{2,0,0}\otimes e_0\rangle
 =-\frac1{2\pi}
   \bigl\||\cdot|^{-1}*(e_0\psi_{2,0,0})\bigr\|_2^2
 =-\frac{448}{19683}.
$$
The second identity follows by integration by parts using $\Delta|x|=2/|x|$ and $-\Delta|x|^{-1}=4\pi\delta_0$. The $\ell=1$ block of $\mathcal P_2|x-y|\mathcal P_2$ is bounded by $\sqrt{35}/2<3$. Rotational invariance and the entries of the $\ell=0$ block computed above give
\begin{equation}\label{eq:appendix-shell-first-moment}
 \mathcal P_2|x-y|\mathcal P_2
 \leq\Bigl(\frac{1007}{324}+\frac{448}{19683}\Bigr)\mathcal P_2
 \leq\frac{47}{15}\mathcal P_2.
\end{equation}

By \eqref{eq:appendix-shell-second-moments}, $\mathcal P_3|x-y|\mathcal P_3\leq(29/4)\mathcal P_3$. Moreover,
\begin{align*}
 \mathcal P_2|x-y|\mathcal P_3|x-y|\mathcal P_2
 &\leq\frac{45}{4}\mathcal P_2
       -\bigl(\mathcal P_2|x-y|\mathcal P_2\bigr)^2\\
 &\leq\Bigl(\frac{144}{5}-\frac{29}{4}\Bigr)
       \Bigl(\frac{16}{5}\mathcal P_2
                    -\mathcal P_2|x-y|\mathcal P_2\Bigr).
\end{align*}
The last inequality follows because the difference of the two scalar quadratics decreases on $[0,47/15]$ and equals $1/225$ at the right endpoint. The block Cauchy--Schwarz inequality therefore gives
$$
 (\mathcal P_2+\mathcal P_3)|x-y|(\mathcal P_2+\mathcal P_3)
 \leq\frac{16}{5}\mathcal P_2+\frac{144}{5}\mathcal P_3.
$$
Weighted Cauchy--Schwarz gives, for $f\in H^1(\R^6)$,
\begin{align*}
 \langle f,\mathbb W f\rangle
 &\geq\sup_{0\neq g\in\Ran(\mathcal P_2+\mathcal P_3)}
       \frac{|\langle g,f\rangle|^2}{\langle g,|x-y|g\rangle}\\
 &\geq\frac5{16}\|\mathcal P_2f\|_2^2
       +\frac5{144}\|\mathcal P_3f\|_2^2.
\end{align*}
Together with the hydrogen spectrum, this implies
\begin{align*}
 H_{2,2}^{\mathrm{sc}}
 &\geq2\lambda_1(P_0\otimes P_0)
       +\Bigl(\lambda_1+\lambda_2+\frac5{16}\Bigr)\mathcal P_2
       +\Bigl(\lambda_1+\lambda_3+\frac5{144}\Bigr)\mathcal P_3\\
 &\qquad
       +(\lambda_1+\lambda_4)
          (\1-P_0\otimes P_0-\mathcal P_2-\mathcal P_3)\\
 &\geq-4(P_0\otimes P_0)
       -\frac{35}{16}(\1-P_0\otimes P_0).
\end{align*}
Since $P_0\otimes P_0$ has rank one, the min--max principle proves \eqref{eq:appendix-spectral-gap}.
\end{proof}

\subsection{Ground-state energy and overlap bounds}\label{app:energy-overlap}

\begin{proposition}\label{prop:appendix-energy-overlap}
We have
\begin{equation}\label{eq:appendix-energy-bounds}
 -\frac{2323}{800}<E^{\F}(2,2)<-\frac{14511}{5000}
\end{equation}
and
\begin{equation}\label{eq:appendix-overlap-bound}
 \|a_\downarrow(e_0)a_\uparrow(e_0)\varphi\|^2
 =|\langle e_0\otimes e_0,\varphi\rangle|^2
 >\frac{181}{200}.
\end{equation}
Consequently, \eqref{eq:core-inputs-main} holds with $\mu=-35/16$ and $\varepsilon_0=19/200$.
\end{proposition}

\begin{proof}
Take $\psi$ to be the normalized positive function proportional to
$$
 e^{-\frac{11}{6}(|x|+|y|)}
 \Bigl(1+\frac3{10}|x-y|
          +\frac2{15}(|x|-|y|)^2\Bigr).
$$
To evaluate its energy and overlap, set $s=|x|+|y|$, $t=|x|-|y|$ and $u=|x-y|$. Spherical integration gives $\dd x\,\dd y=\pi^2u(s^2-t^2)\dd s\,\dd t\,\dd u$ on $0<u<s$, $-u<t<u$, and, for integers $i,j,k\geq0$ and $\beta>0$,
$$
 \int_0^\infty e^{-\beta s}\int_0^s\int_{-u}^u
 s^iu^jt^{2k}\dd t\,\dd u\,\dd s
 =\frac{2(i+j+2k+2)!}
 {(2k+1)(j+2k+2)\beta^{i+j+2k+3}}.
$$
Combining the ground-state transform with $e^{-11s/6}$ and the integral identity gives
$$
 E^{\F}(2,2)\leq\langle\psi,H_{2,2}^{\mathrm{sc}}\psi\rangle
 =-\frac{158492389}{54610893}
 <-\frac{14511}{5000},
 \qquad
 |\langle e_0\otimes e_0,\psi\rangle|^2>\frac{67}{72}.
$$
The lower energy bound follows from Bazley's rigorous estimate $E^{\F}(2,2)\geq-2.9037474$ \cite{Bazley60}.

Using \eqref{eq:appendix-spectral-gap}, we obtain
\begin{align*}
 1-|\langle\varphi,\psi\rangle|^2
 &\leq
 \frac{\langle\psi,H_{2,2}^{\mathrm{sc}}\psi\rangle-E^{\F}(2,2)}
      {-35/16-E^{\F}(2,2)}
 <\frac{31}{14325}.
\end{align*}
Here the quotient decreases as $E^{\F}(2,2)$ increases. Orthogonal decomposition along $\psi$ and Cauchy--Schwarz now give
\begin{align*}
 \langle e_0\otimes e_0,\varphi\rangle
 &\geq
 \langle e_0\otimes e_0,\psi\rangle
 \langle\psi,\varphi\rangle-
 \sqrt{1-|\langle e_0\otimes e_0,\psi\rangle|^2}\,
 \sqrt{1-|\langle\psi,\varphi\rangle|^2}\\
 &>
 \sqrt{\frac{67}{72}\Bigl(1-\frac{31}{14325}\Bigr)}
 -\sqrt{\frac5{72}\frac{31}{14325}}
 >\sqrt{\frac{181}{200}}.
\end{align*}
\end{proof}

\subsection{Density moments}\label{app:source-moments}

Throughout this subsection we take $\mu=-35/16$ and $\varepsilon_0=19/200$, using Proposition~\ref{prop:appendix-energy-overlap}.

\begin{proposition}\label{prop:appendix-density-moments}
We have
$$
\begin{gathered}
\operatorname{Tr}(|\cdot|^2\gamma_\phi)\le\frac{13}{10},
\qquad
\operatorname{Tr}(|\cdot|^2\Gamma_{+,\phi})\le\frac{41}{500},
\qquad
\operatorname{Tr}\Gamma_{+,\phi}\le\frac1{16},
\\
\mu-E^{\F}(2,2)\ge\frac57,
\qquad
\langle e_0,|\cdot|^2e_0\rangle=\frac34,
\qquad
\|Q_0|\cdot|^2e_0\|^2=\frac{27}{32}.
\end{gathered}
$$
\end{proposition}

\begin{proof}
Proposition~\ref{prop:appendix-energy-overlap} gives $\mu-E^{\F}(2,2)>5/7$. Direct integration with $e_0(x)=\sqrt{8/\pi}\,e^{-2|x|}$ gives $\langle e_0,|\cdot|^k e_0\rangle=(k+2)!/(2\cdot4^k)$ for $k=2,4$, proving the last two identities.

For the normalized trial function $\psi$ in the proof of Proposition~\ref{prop:appendix-energy-overlap}, spherical integration gives
$$
\begin{array}{ll}
 \||x|\psi\|_2^2<\dfrac{29}{25},
 &\displaystyle
 \left\|\left(\frac{|x|^2+|y|^2}{2}-\frac{29}{25}\right)
             \psi\right\|_2<\frac{21}{20},\\[2mm]
 \|(\1\otimes Q_0)\psi\|_2<\dfrac15,
 &\||x|(\1\otimes Q_0)\psi\|_2^2<\dfrac{23}{1000},\\[2mm]
 \||x|^2(\1\otimes Q_0)\psi\|_2<\dfrac15,
 &\||x|^2(H_{2,2}^{\mathrm{sc}}-E^{\F}(2,2))\psi\|_2<\dfrac12.
\end{array}
$$
The last estimate holds throughout \eqref{eq:appendix-energy-bounds}. Decompose $\phi=\langle\psi,\phi\rangle\psi+\chi$, with $\chi\perp\psi$. The preceding proof also gives
$$
 \|\chi\|_2^2
 =1-|\langle\psi,\phi\rangle|^2
 <\frac{31}{14325}<\frac1{441}.
$$
Rotational invariance implies $x_j\chi\perp\phi$. Thus the spectral gap and the ground-state equation, justified with radial cutoffs, give
\begin{align*}
 (\mu-E^{\F}(2,2))\||x|\chi\|_2^2
 &\leq\sum_{j=1}^3
   \langle x_j\chi,(H_{2,2}^{\mathrm{sc}}-E^{\F}(2,2))x_j\chi\rangle\\
 &=-\langle\psi,\phi\rangle\operatorname{Re}
   \langle\chi,|x|^2(H_{2,2}^{\mathrm{sc}}-E^{\F}(2,2))\psi\rangle
   +\frac32\|\chi\|_2^2
 <\frac1{42}+\frac1{294}.
\end{align*}
Consequently, $\||x|\chi\|_2^2<4/105$. Expanding with the same decomposition and using exchange symmetry and $\chi\perp\psi$, we obtain
\begin{align*}
 \operatorname{Tr}(|\cdot|^2\gamma_\phi)
 &=\||x|\phi\|_2^2\\
 &\leq\||x|\psi\|_2^2
   +2\|\chi\|_2
    \left\|\left(\frac{|x|^2+|y|^2}{2}-\frac{29}{25}\right)
                \psi\right\|_2
   +\||x|\chi\|_2^2\\
 &<\frac{29}{25}+\frac1{10}+\frac4{105}<\frac{13}{10}.
\end{align*}
Likewise, since $\1\otimes Q_0$ commutes with $|x|$,
\begin{align*}
 \operatorname{Tr}(|\cdot|^2\Gamma_{+,\phi})
 &=\||x|(\1\otimes Q_0)\phi\|_2^2\\
 &\leq\||x|(\1\otimes Q_0)\psi\|_2^2
   +2\|\chi\|_2\,\||x|^2(\1\otimes Q_0)\psi\|_2
   +\||x|\chi\|_2^2\\
 &<\frac{23}{1000}+\frac2{105}+\frac4{105}<\frac{41}{500}.
\end{align*}
Finally,
$$
 (\operatorname{Tr}\Gamma_{+,\phi})^{1/2}
 =\|(\1\otimes Q_0)\phi\|_2
 \leq\|(\1\otimes Q_0)\psi\|_2+\|\chi\|_2
 <\frac15+\frac1{21}<\frac14.
$$
\end{proof}


\begin{thebibliography}{99}

\bibitem{Bazley60}
N. W. Bazley, \emph{Lower bounds for eigenvalues with application to the
helium atom}, Phys. Rev. \textbf{120} (1960), 144--149.
\href{https://doi.org/10.1103/PhysRev.120.144}{doi:10.1103/PhysRev.120.144}.

\bibitem{BL85}
R. Benguria and E. H. Lieb,
\emph{The most negative ion in the Thomas--Fermi--von Weizs\"acker
theory of atoms and molecules},
J. Phys. B: At. Mol. Phys. \textbf{18} (1985), 1045--1059.
\href{https://doi.org/10.1088/0022-3700/18/6/006}{doi:10.1088/0022-3700/18/6/006}.

\bibitem{ChenWu26}
L. Chen and Y. Wu,
\emph{Asymptotic neutrality for the ionization conjecture in the
two-center Thomas--Fermi--Dirac--von Weizs\"acker theory},
preprint, \href{https://arxiv.org/abs/2610.09997}{arXiv:2610.09997} (2026).

\bibitem{FS90}
C. L. Fefferman and L. A. Seco,
\emph{Asymptotic neutrality of large ions},
Comm. Math. Phys. \textbf{128} (1990), 109--130.
\href{https://doi.org/10.1007/BF02097048}{doi:10.1007/BF02097048}.

\bibitem{FK08}
D. M. Feldmann and F. W. King,
\emph{Upper bound to the critical binding nuclear charge for a
three-electron atomic system},
J. Phys. B: At. Mol. Opt. Phys. \textbf{41} (2008), 025002.
\href{https://doi.org/10.1088/0953-4075/41/2/025002}{doi:10.1088/0953-4075/41/2/025002}.

\bibitem{FNV18}
R. L. Frank, P. T. Nam, and H. Van Den Bosch,
\emph{The ionization conjecture in Thomas--Fermi--Dirac--von
Weizs\"acker theory},
Comm. Pure Appl. Math. \textbf{71} (2018), 577--614.
\href{https://doi.org/10.1002/cpa.21717}{doi:10.1002/cpa.21717}.

\bibitem{GST96}
F. Gesztesy, B. Simon, and G. Teschl,
\emph{Zeros of the Wronskian and renormalized oscillation theory},
Amer. J. Math. \textbf{118} (1996), 571--594.
\href{https://doi.org/10.1353/ajm.1996.0024}{doi:10.1353/ajm.1996.0024}.

\bibitem{Hogreve11}
H. Hogreve,
\emph{A remark on the ground state energy of bosonic atoms},
J. Stat. Phys. \textbf{144} (2011), 904--908.
\href{https://doi.org/10.1007/s10955-011-0265-0}
{doi:10.1007/s10955-011-0265-0}.

\bibitem{HPS25}
D. Hundertmark, N. Pattakos, and M. R. Schulz,
\emph{On the excess charge problem of atoms},
preprint, \href{https://arxiv.org/abs/2504.18487}{arXiv:2504.18487} (2025).

\bibitem{KNS99}
S. Kais, J. P. Neirotti, and P. Serra,
\emph{Phase transitions and the stability of atomic and molecular ions},
Int. J. Mass Spectrom. \textbf{182--183} (1999), 23--29.
\href{https://doi.org/10.1016/S1387-3806(98)14224-0}
{doi:10.1016/S1387-3806(98)14224-0}.

\bibitem{Lewin25}
M. Lewin,
\emph{Some open mathematical problems concerning charged quantum particles},
C. R. Phys. \textbf{26} (2025), 369--380.
\href{https://doi.org/10.5802/crphys.249}{doi:10.5802/crphys.249}.

\bibitem{Lieb84}
E. H. Lieb,
\emph{Atomic and molecular negative ions},
Phys. Rev. Lett. \textbf{52} (1984), 315--317.
\href{https://doi.org/10.1103/PhysRevLett.52.315}{doi:10.1103/PhysRevLett.52.315}.

\bibitem{LiebSeiringer10}
E. H. Lieb and R. Seiringer,
\emph{The Stability of Matter in Quantum Mechanics},
Cambridge University Press, Cambridge, 2010.
\href{https://doi.org/10.1017/CBO9780511819681}
{doi:10.1017/CBO9780511819681}.

\bibitem{LSST88}
E. H. Lieb, I. M. Sigal, B. Simon, and W. Thirring,
\emph{Approximate neutrality of large-$Z$ ions},
Comm. Math. Phys. \textbf{116} (1988), 635--644.
\href{https://doi.org/10.1007/BF01224904}{doi:10.1007/BF01224904}.

\bibitem{Nam12}
P. T. Nam, \emph{New bounds on the maximum ionization of atoms},
Comm. Math. Phys. \textbf{312} (2012), 427--445.
\href{https://doi.org/10.1007/s00220-012-1479-y}{doi:10.1007/s00220-012-1479-y}.

\bibitem{Nam22}
P. T. Nam, \emph{The ionization problem in quantum mechanics},
in \emph{The Physics and Mathematics of Elliott Lieb:
The 90th Anniversary}, Vol.~II,
R. L. Frank, A. Laptev, M. Lewin, and R. Seiringer (eds.),
EMS Press, Berlin, 2022, pp.~93--120.
\href{https://doi.org/10.4171/90-2/34}{doi:10.4171/90-2/34}.

\bibitem{OpenAI26}
OpenAI, \emph{Uniform excess charge for Coulomb molecules and the
outer radius of neutral atoms},
OpenAI Math Release preprint, September~24, 2026.
\href{https://github.com/openai/math/blob/main/preprints/Uniform-excess-charge-for-Coulomb-molecules-and-the-outer-radius-of-neutral-atoms-September-24-2026/paper.pdf}{Preprint}.

\bibitem{SSS90}
L. A. Seco, I. M. Sigal, and J. P. Solovej,
\emph{Bound on the ionization energy of large atoms},
Comm. Math. Phys. \textbf{131} (1990), 307--315.
\href{https://doi.org/10.1007/BF02161416}{doi:10.1007/BF02161416}.

\bibitem{SYT26}
A. Shelley, C. F. Yang, and D. M. Tartakovsky,
\emph{Sharp ionization bounds for helium and lithium},
Hexagon preprint \href{https://hexagonmath.org/2611.00325v1}{2611.00325v1} (2026).

\bibitem{Solovej90}
J. P. Solovej, \emph{Asymptotics for bosonic atoms},
Lett. Math. Phys. \textbf{20} (1990), 165--172.
\href{https://doi.org/10.1007/BF00398282}{doi:10.1007/BF00398282}.

\bibitem{Solovej03}
J. P. Solovej, \emph{The ionization conjecture in Hartree--Fock theory},
Ann. of Math. (2) \textbf{158} (2003), 509--576.
\href{https://doi.org/10.4007/annals.2003.158.509}{doi:10.4007/annals.2003.158.509}.

\bibitem{SS74}
F. H. Stillinger and D. K. Stillinger,
\emph{Nonlinear variational study of perturbation theory for atoms and ions},
Phys. Rev. A \textbf{10} (1974), 1109--1121.
\href{https://doi.org/10.1103/PhysRevA.10.1109}{doi:10.1103/PhysRevA.10.1109}.

\bibitem{Teschl09}
G. Teschl, \emph{Mathematical Methods in Quantum Mechanics: With Applications
to Schr\"odinger Operators}, Graduate Studies in Mathematics \textbf{99},
American Mathematical Society, Providence, RI, 2009.
\href{https://doi.org/10.1090/gsm/099}{doi:10.1090/gsm/099}.
\href{https://www.mat.univie.ac.at/~gerald/ftp/book-schroe/schroe.pdf}{Author's online edition}.

\bibitem{WLYD14}
L. Wang, C. Li, Z.-C. Yan, and G. W. F. Drake,
\emph{Fine structure and ionization energy of the $1s2s2p$ ${}^{4}P$
state of the helium negative ion $\mathrm{He}^{-}$},
Phys. Rev. Lett. \textbf{113} (2014), 263007.
\href{https://doi.org/10.1103/PhysRevLett.113.263007}
{doi:10.1103/PhysRevLett.113.263007}.

\bibitem{Zhislin60}
G. M. Zhislin,
\emph{A study of the spectrum of the Schr\"odinger operator for a
system of several particles},
Tr. Mosk. Mat. Obs. \textbf{9} (1960), 81--120 (Russian).
\url{https://www.mathnet.ru/eng/mmo97}.
\href{https://www.mathnet.ru/eng/mmo127}{Erratum}, Tr. Mosk. Mat. Obs. \textbf{10} (1961), 437.

\bibitem{ZhislinSigalov65}
G. M. Zhislin and A. G. Sigalov,
\emph{The spectrum of the energy operator for atoms with fixed nuclei
on subspaces corresponding to irreducible representations of the group
of permutations},
Izv. Akad. Nauk SSSR Ser. Mat. \textbf{29} (1965), 835--860 (Russian).
\url{https://www.mathnet.ru/eng/im3076}.

\end{thebibliography}
\end{document}